{}~~~~~~~~~~~~~~~~~~~~~~~~~~~~~~~~~~~~~~~~~~~~~~~~~~~~~~~~~~~~~~~

\documentstyle[preprint,eqsecnum,aps]{revtex}

\begin{document}
\draft
\preprint{Phys. Rev. D (in printing)}

\title{Model Calculation of Nucleon Structure Functions}

\author{X. Song and J. S. McCarthy}

\address{Institute of Nuclear and Particle Physics,
Department of Physics,\\ University of Virginia,
Charlottesville, VA 22901, U.S.A.}

\maketitle
\begin{abstract}
Deep inelastic polarized and unpolarized structure
functions for a free nucleon are obtained in a
modified Center-of-Mass bag model, which includes
the symmetry breaking effects from spin-dependent
interactions. The quark distribution functions,
calculated at $Q_0^2\simeq$ (0.9GeV/c)$^2$, are
evolved to higher $Q^2$ region and compared with
the data and other models. The model gives a
reasonable description for the valence part of
the structure functions at $x>0.3$. For small-$x$
region, the contributions from the sea are necessary.
The spin-dependent effects are important
in describing the existing data.
\end{abstract}
\bigskip
\bigskip

\widetext

\section{INTRODUCTION}

It is well known that quantum chromodynamics (QCD), the
theory of interactions of quarks and gluons, provides a
basic description of strong interactions in the standard
model. For high energy scattering processes, the cross
section can be factorized into a $hard$ piece, the parton
cross section which can be calculated in the framework of
perturbative QCD, and a $soft$ piece which depends on the
parton distribution functions inside the nucleon. In the
QCD quark parton model, at the leading order, the structure
functions in the deep inelastic lepton-nucleon scattering
are `charge' squared weighted combinations of the quark
distribution functions $q_{f,h}(x)$, which denotes the
probability finding a quark with flavor $f$, momentum
fraction $x$ and helicity $h$ within the nucleon. These
distribution functions are essentially determined by the
quark-gluon structure of the nucleon. Since these functions
play an important role in standard model phenomenology
and in understanding the nucleon structure, many recent
experiments have been done to measure deep inelastic
unpolarized structure functions and extract the parton
distribution functions. (see \cite{sloan88,mishra89} for
the review of deep inelastic structure functions,
\cite{roberts91} for a comprehensive compilation of the
latest deep inelastic data and \cite{owens92,milsztajn91}
for the comparisons of data from different measurements.)

Among the unpolarized data, recent deep inelastic
scattering data provide some new information of the
internal structure of the nucleon. In particular,
from the recent NMC measurement \cite{amaudruz91} of
difference $F_2^p-F_2^n$ the Gottfried sum rule
\cite{gottfried67} is found to be significantly less
than the quark-parton model prediction of 1/3. This
implies that down-sea is larger than up-sea. Combining
with a smaller strange-sea from neutrino scattering
\cite{abram82,foudas90}, it seems likely that the
unpolarized sea is not only SU(3) flavor asymmetric,
but also SU(2) flavor asymmetric (i.e. down-sea and
up-sea are not equal). However, an SU(2) flavor
symmetric sea is usually assumed.

For the polarized structure functions, many experiments
\cite{alguard76,baum80,ashman89} have been performed in
the past two decades, The review of earlier works can
be found in \cite{hughes83}. The recent measurement
from EMC at CERN shows \cite{ashman89} that the first
moment of $g_1^p$ is significantly below the value
expected from the Ellis-Jaffe sum rule \cite{ellis74}.
Furthermore, the EMC data combined with the Bjorken
sum rule \cite{bjorken66} and the hyperon Beta-decay
data indicates that only a small fraction of the
proton's spin is carried by the spin angular momentum
of the quarks, which seems to contradict the naive
expectations from the low energy constituent quark
models. These conclusions are confirmed by two most
recent experimental results reported by SMC \cite{adeva93}
and E142 \cite{henley93} within large errors. At
first glance the E142 data seems to disagree with
the Bjorken sum rule, but a detail analysis given
by Ellis and Karliner
\cite{ek93} shows that the discrepancy between the
conclusions drawn by the SMC and E142 data can be
eliminated if the $Q^2$-variation and higher twist
effects are taken into account.

There are a variety of theoretical activities
attempting to resolve the quantitative decomposition
and the origin of the spin of the nucleon.
An incomplete list of recent works and review on this
topic see \cite{brodsky88,altarelli88,carlitz88,efremov88}
and \cite{altarelli89,jaffe90,reya91,griffioen92}, and
references there in. In principle, one should be able to
calculate the parton distribution functions from the
basic equations of QCD which determine the quark-gluon
structure of the nucleon. It needs to know the basic
constituents inside the nucleon and QCD (perturbative
and nonperturbative) interactions among these
constituents, hence the structure of the nucleon in
QCD is remarkably complex. The lattice QCD \cite{ukawa93}
has provided a framework of evaluating the hadron structure
(masses, quark antiquark potential and other low energy
observables) in the nonperturbative way. There have been
several preliminary results \cite{altmeyer93,dong93} of
the quark spin fraction of the nucleon spin have been
reported. In \cite{dong93}, the quark loop (or `sea' quark)
contributions are calculated. Even so, the evaluation of
the structure functions and parton distribution functions
is still not accessible in lattice approach and has
to resort to various QCD inspired nucleon models.

The first pioneering work of using MIT bag model to
calculate deep inelastic (unpolarized) structure
functions was made by Jaffe \cite{jaffe75}, then the
polarized structure functions were calculated in the
same model by many authors( Hughes \cite{hughes77},
Bell \cite{bell78} and Celenza and Shakin \cite{celenza83} )
and recently by Jaffe and Ji \cite{jaffe91}. Since the
momentum is not conserved in the intermediate state in
the MIT bag model the structure function does not
vanish for $x\geq 1$. An attempt to remove
this difficulty by restoring the four momentum
conservation and including the effects of spectator
quarks was given in the CM (center-of-mass) bag
model \cite{wang83}. The main idea of the model is that
instead of the single quark current in the MIT bag
model a covariant effective electromagnetic current
of the nucleon was suggested. The current includes
not only the piece of the struck quark but also the
pieces of the spectator quarks. The current satisfies
translational invariance so four-momentum is
conserved. Another approach to avoid the support
problem in the MIT bag model was suggested by
Benesh and Miller \cite{benesh87}. The Peierls-Yoccoz
projection \cite{peierls57} has been used to obtain an
eigenstate of zero momentum. Including pion
corrections the valence structure functions of
the nucleon and $\Delta$ are calculated. Another
disadvantage of the original MIT bag model is that
the spin-dependent force effect, which has been
used to successfully explain the hadron spectroscopy
\cite{rujula75}, was neglected, hence
$F_2^n(x)/F_2^p(x)$=2/3 is a constant in
disagreement with data. Many models
\cite{meyer91,schreiber91,bate92,bickerstaff90,traini92}
have been suggested to incorporate the spin-dependent
effects in describing the inelastic structure functions.
In Ref. \cite{meyer91}, the nucleon is considered as a
composite system of a quark and a pointlike diquark.
The model treat kinematics relativistically and
considers the spin-flavor dependence in the nucleon
wave function. In Ref. \cite{schreiber91}, the effect
of one-gluon exchange, yielding the $N-\Delta$ mass
difference, is incorporated and significantly modifies
the result given in the SU(6)-symmetric model. In
\cite{bate92}, the structure functions are calculated
by using two phenomenological nucleon models $-$
non-topological soliton (NTS) model and
color-dielectric (CD) model. For NTS model, the
authors conclude that an agreement between $xg_1^p$
and the EMC data could be obtained even the proton
wave function is SU(6)-symmetric. As the authors of
\cite{meyer91} pointed out, however, that the model
they considered can not reproduce the large body of
low energy data. In Ref. \cite{bickerstaff90}, the
nucleon, which consists of three confining quarks,
is treated approximately as an infinite free Fermi
gas system at finite temperature. In \cite{traini92},
the constituent quark model \cite{isgur78} of the
nucleon are used. In both \cite{bickerstaff90} and
\cite{traini92}, only unpolarized structure functions
are calculated. Most of these model results are fairly
close and can compared with the data in spite of
different assumptions used in different models. It
should be noted that a common features of all these
calculations is that the structure functions are
evaluated at some very low momentum transfer scale
${Q_0}^2$ (e.g $\sim 0.063 $ GeV$^2$ for Ref.
\cite{meyer91} and \cite{traini92}, $\sim 0.068$
GeV$^2$ for Ref. \cite{schreiber91}, $\sim 0.09$
GeV$^2$ for Ref. \cite{bate92}, and $\sim 0.06$
GeV$^2$ for Ref. \cite{bickerstaff90} ), then these
functions are evolved to the higher $Q^2$ region by
using leading order perturbative QCD. The problem,
as pointed out by the authors of Ref. \cite{meyer91},
is that `in order to obtain the correct momentum sum
rule, it is necessary to go to extreme value of
strong coupling constant ${\alpha}_s\geq 3-4$.
Therefore the use of QCD perturbation theory at the
leading order can not be justified'. We will show,
however, that one may avoid this ambiguity and still
obtain a reasonable description of deep inelastic
structure functions from the low energy nucleon model.

Recently, we suggested a modified CM bag model (MCM)
\cite{song92}, which based on the CM bag model in
\cite{wang83}, to calculate
both elastic form factors and deep inelastic structure
functions. In \cite{song92}, the symmetry breaking
effects coming from perturbative one-gluon exchange
mechanism (color hyperfine interactions) and/or
nonperturbative instanton spin interactions are
simulated by introducing a symmetry breaking parameter.
It has been shown that by including spin force effects,
the model gives a fairly good description for the
magnetic moments of octet baryon and the electromagnetic
form factors of the nucleon. We also shown briefly in
\cite{song92} that the same model can fairly describe
many aspects of deep inelastic structure functions of
the nucleon. In this paper, we present detail results
of these calculations.

The paper is organized as follows: in section II, we
will define the notations and give the basic formalism
of the model calculation of the structure functions,
and list the approximations used in the model. In
section III, the unpolarized structure functions $F_1$
and $F_2$ are derived from the hadronic tensor. In the
Bjorken limit, these functions are scaling and satisfy
the Callen-Gross relation. The valence
quark distribution functions $xu_v(x,Q_0^2)$,
$xd_v(x,Q_0^2)$ are evaluated around
$Q_0^2\simeq$(0.9 GeV/c)$^2$ from the model and then
evolved to $Q^2=4$, 10.7 and 15 (GeV/c)$^2$ by using
leading order perturbative QCD evolution equations
\cite{altarelli77} and the analytical approach
developed in \cite{shen83}. Including the sea
contributions, the structure functions $F_2^{p}$,
$F_2^{n}$, $F_2^{p}-F_2^{n}$, the integral
$\int_0^1dx[F_2^{p}-F_2^{n}]_{val}/x$ and the ratio
$F_2^{n}/F_2^{p}$ are compared with data and those
from other models. In section IV, the spin dependent
structure functions $g_1$ and $g_2$ are derived in
the model. The polarized quark distributions,
$\Delta u_v(x,Q^2)$, $\Delta d_v(x,Q^2)$ and their
first moments are evaluated in the model at the same
$Q_0^2$ and then evolved to higher $Q^2$ region
\cite{song89}. The polarized sea
quark distributions are discussed. Using the Bjorken
sum rule, EMC data, $\nu$-p scattering data and the
model results of $\Delta u_v$ and $\Delta d_v$, a
set of sea quark polarizations is obtained. Including
the sea contribution, the $xg_1^{(p)}(x)$,
$xg_1^{(n)}(x)$ and $g_1^d(x)$ are compared with the
latest data \cite{ashman89,adeva93,henley93}. In
section V, the second spin dependent structure function
$g_2(x)$, the Burkhardt-Cottingham sum rule
\cite{burkhardt70} and the higher twist effects
are briefly discussed. A summary is drawn in section VI.
Derivations of some formulae are given in Appendix
I and II.

\section{KINEMATICS AND HADRONIC TENSOR}

In the one photon exchange approximation, the deep
inelastic structure functions can be extracted from
the hadronic tensor, which is the Fourier
transformation of the single nucleon matrix
element for the commutator of two electromagnetic
currents

\begin{equation}
$$W_{\mu\nu}(P,q,S)={1\over {4\pi}}\int d^4ye^{iqy}<P,S\mid
[J_{\mu}(y), J_{\nu}(0)]\mid P,S>               $$
\end{equation}
where $J_{\mu}(y)$ is the hadronic electromagnetic
current which depends on the nucleon model one used,
$P^{\mu}$ and $S^{\mu}$ are the four-momentum and spin
vector of the target nucleon ($P^{\mu}P_{\mu}=M^2$,\
$M$ is nucleon mass;\ $S^{\mu}S_{\mu}=-1$ and
$P^{\mu}S_{\mu}=0$) respectively.
The initial nucleon state is covariantly normalized:
$<P,S\mid P',S'>=({2\pi})^32P^0{\delta}^3({\bf P}-{\bf P'})
{\delta}_{SS'}$ and $q^{\mu}$ is the four-momentum of
the virtual-photon. The conventional kinematical variables
$Q^2$ and $\nu$ (or $Q^2$ and $x$) are defined as:
$Q^2\equiv -q^2>0$ and $x\equiv Q^2/2M\nu$ ($\nu=
P\cdot q/M$) with $0\leq x\leq 1$.

$W_{\mu\nu}$ in (2.1) can be decomposed into two
parts
\begin{equation}
$$W_{\mu\nu}=W_{\mu\nu}^{(S)}+iW_{\mu\nu}^{(A)}  $$
\end{equation}
where
\begin{equation}
$$W_{\mu\nu}^{(S)}=(-g_{\mu\nu}+{q_{\mu}q_{\nu}}/{q^2})
W_1(x, Q^2)+(P_{\mu}-q_{\mu}{P\cdot q}/{q^2})
(P_{\nu}-q_{\nu}{P\cdot q}/{q^2}){W_2(x, Q^2)}/{M^2}
$$
\end{equation}
and
\begin{equation}
$$W_{\mu\nu}^{(A)}={\epsilon}_{\mu\nu\lambda\sigma}
 q^{\lambda}\{ S^{\sigma}MG_1(x, Q^2)+[(P\cdot q)S^{\sigma}
-({q\cdot S})P^{\sigma}]{G_2(x, Q^2)}/{M} \}
                                                 $$
\end{equation}
represent, respectively, the symmetric and antisymmetric
parts of the tensor, from which the structure functions $F_1$,
$F_2$ and $G_1$, $G_2$ are defined. In the
deep inelastic region, these structure functions
become the scaling functions of the Bjorken
variable $x$ only, i.e. in the Bjorken limit
($Q^2\rightarrow +\infty$, $\nu \rightarrow +\infty$
with $x$ fixed)

\begin{equation}
$$MW_1(x,Q^2)\rightarrow F_1(x)\ ;\quad
\nu W_2(x,Q^2)\rightarrow F_2(x)                  $$
\end{equation}
and

\begin{equation}
$$M^2{\nu}G_1(x,Q^2)\rightarrow g_1(x)\ ;\quad
M{\nu}^2G_2(x,Q^2)\rightarrow g_2(x)          $$
\end{equation}
The main purpose of this paper is to calculate
these structure functions by using the current $J_{\mu}$
given in the modified CM bag model. The assumptions or
approximations we used are listed below:

($\bf a$) The effective nucleon current is the sum of
single quark current, i.e. the virtual photon interacts
with only one quark (struck quark) at a time and other
two quarks are spectators, this is the impulse approximation.

($\bf b$) The nucleon is assumed to be in the Fock state
which only consists of three valence quarks. $\bf 56$-plet
SU(6) wave functions for the proton and neutron with
spin-up are used, the symmetry breaking effect is
described in terms of a parameter \cite{song92}
$\xi\equiv R_d^p/R_u^p<1$, which simulates the smaller
spatial size for the scalar $u-d$ quark pair than that
for the vector $u-u$ or $d-d$ quark pairs in the nucleon.

($\bf c$) The effect of quark confinement is described
in terms of a bound state quark spatial wave function,
which is basically determined by the large-scale structure
of the nucleon.

Based on approximations ({\bf a}) and ({\bf b}), the
$\gamma$NN vertex can be written \cite{song92}

\begin{eqnarray}
\int d^4ye^{i{q}\cdot{y}}<p'\mid J_{\mu}(y)\mid p>
&=&(2{\pi})^4{\delta}^4(p+q-p')<p'\mid J_{\mu}(0)\mid p>\nonumber \\
&=&(2{\pi}){\delta}^4(p+q-p')\sum\limits_{{\bf 1}\rightarrow {\bf 2,3}}
\int \prod\limits_{i={\bf 1}}^{\bf 3}d^3{\bf r_i}e^{i{\bf q}
\cdot{\bf r}_1}\nonumber \\
&\cdot& {\bar q}_{p',{\alpha}'}({\bf r}_1,{\bf r}_2,{\bf r}_3)
[{\hat e}_q{\gamma}_{\mu}]_{\bf 1}
q_{p,{\alpha}}({\bf r}_1,{\bf r}_2,{\bf r}_3)
\end{eqnarray}
where the $\delta$-function comes from the center-of-mass
motion. The integral on the r.h.s. of (2.7) denotes the
contribution coming from the $\gamma qq$ interaction and
quark internal motion, where subscript $\bf 1$ of the
operator $[{\hat e}_q{\gamma}_{\mu}]_{\bf 1}$ denotes
that the virtual photon interacts with the first (struck)
quark and subscripts $\bf 2$ and $\bf 3$ denote the
spectators. The summation runs over three quarks (${\bf 1}
\rightarrow {\bf 3}$). The nucleon wave function in (2.7)
is

\begin{equation}
$$q_{p,{\alpha}}({\bf r}_1,{\bf r}_2,{\bf r}_3)=
\prod\limits_{i=1}^3 q_{p,m}({\bf r}_i){\alpha}_N
$$
\end{equation}
where ${\alpha}_N$ is the SU(6) spin-flavor wave
function of the nucleon, and the antisymmetric color
wave function has been omitted. In (2.8),
$q_{p,m}({\bf r}_i)$ are the spin$-1/2$
bound quark wave functions. To calculate the
Lorentz tensor $W_{\mu\nu}$, we use the
nucleon rest frame. In this frame only the rest
frame wave functions $q_m({\bf r}_i)$ are involved,
they can be described by, for instance,
the cavity solution in the MIT bag model \cite{jaffe75}
or other bound quark wave function in the
relativistic quark model (see (9), (10), (19) and (20)
in Appendix I).

Using the current (2.7), the hadronic tensor (2.1)
can be written as

\begin{eqnarray}
W_{\mu\nu}(P,q,S)&=&
\sum\limits_{{\bf 1}\rightarrow {\bf 2,3}}
\sum\limits_{{\alpha}_1,m_1}b_{{\alpha}_1,m_1}({\bf 1;23})
{M\over {(2{\pi})^6{R}_1^3}}
\int\prod\limits_{i={\bf 1}}^{\bf 3}{{d^3{\bf k}_i}\over {2k_i}}\nonumber \\
&&\cdot {\delta}^4(q+P-{\sum\limits_ik_i})
I_{m_1\mu\nu}({\bf k_1-q})
I_{m_2}({\bf k_2})I_{m_3}({\bf k}_3)
\end{eqnarray}
where $b_{{\alpha}_1,m_1}({\bf 1};{\bf 23})
=<\alpha_N\mid [b^{\dagger}_{\alpha}(m)
{\hat e_q}^2b_{\alpha}(m)]_{\bf 1}\mid \alpha_N >$
is the matrix element of ${\hat e}_q^2$ (the charge
square operator of the struck quark),
${\bf k}_i$ and $m_i$ are the three-momentum and
spin projection of i-th quark, and $R_i$ (i=1,2,3)
are the parameters appeared in the wave functions of
the struck quark (i=1) and spectator quarks (i=2,3)
which determine the radius of the quark
distributions (see Appendix I). The integral

\begin{equation}
$$I_{m_1,\mu\nu}({\bf k_1-q})\equiv \int
d^3{\bf r}_1\int d^3{\bf r_1}'e^{i({\bf k_1-q})
\cdot ({\bf {r_1}-{r_1}'})}{\bar q}_{m_1}({\bf r}_1)
{\gamma}_{\mu}{\rlap/{k_1}}{\gamma}_{\nu}q_{m_1}({\bf r_1}')
 $$
\end{equation}
denotes the contribution coming from the struck quark
which interacts with the virtual photon which carries
four-momentum $q=(q_0,{\bf q})$ while

\begin{equation}
$$I_{{m}_j}({\bf k_j})\equiv \int d^3{\bf r}_j\int
d^3{\bf r_j}'e^{i{\bf k_j}\cdot ({\bf {r_j}-{r_j}'})}
{\bar q}_{m_j}({\bf r}_j){\gamma}_0{\rlap/{k_j}}
{\gamma}_0q_{m_j}({\bf r_j}')\qquad (j=2,3)
$$
\end{equation}
denote the contributions from the spectator quarks.
The $\delta-$function in (2.9) guarantees the four-momentum
conservation.

Using the identity
${\gamma}_{\mu}{\gamma}_{\lambda}{\gamma}_{\nu}$=
$S_{\mu\lambda\nu\sigma}{\gamma}^{\sigma}
+i{\epsilon}_{\mu\nu\lambda\sigma}{\gamma}^{\sigma}{\gamma}^5$,
where $S_{\mu\lambda\nu\sigma}=g_{\mu\lambda}g_{\nu\sigma}
+g_{\mu\sigma}g_{\nu\lambda}-g_{\mu\nu}g_{\lambda\sigma}$,
the integral (2.11) can be rewritten as

\begin{equation}
$$I_{m_1,\mu\nu}({\bf k_1-q})=S_{\mu\lambda\nu\sigma}
{k_1}^{\lambda}I^{(s)\sigma}({\bf k_1-q})
-i{\epsilon}_{\mu\nu\lambda\sigma}
{k_1}^{\lambda}I_{m_1}^{(A)\sigma}({\bf k_1-q})
$$
\end{equation}
in which the symmetric term $I^{(s)\sigma}({\bf k_1-q})$
is independent of the quark spin projection and contributes to the
unpolarized structure functions $F_{1}(x)$ and $F_{2}(x)$,
while the antisymmetric term $I_{m_1}^{(A)\sigma}({\bf k_1-q})$
depends on the quark spin projection and generates the polarized
structure functions $g_{1}(x)$ and $g_2(x)$. We will discuss them
separately.

\section{UNPOLARIZED STRUCTURE FUNCTIONS}

\subsection{Formalism}

Using standard projection operators and (2.9), we obtain $W_1$
and $W_2$ as follows (see Appendix I)

\begin{eqnarray}
W_{i}&=&
\sum\limits_{{\bf 1}\rightarrow {\bf 2,3}}
\sum\limits_{{\alpha}_1,m_1}b_{{\alpha}_1,m_1}({\bf 1; 23})
{M\over {(2{\pi})^6{R}_1^3}}
\int\prod_{j={\bf 1}}^{\bf 3}{{d^3{\bf k}_j}\over {2k_j}}
{\delta}^4(q+P-{\sum\limits_jk_j})\nonumber \\
\qquad &\cdot& I_{m_1}^{(i)}
({\bf k_1-q})I_{m_2}({\bf k_2})I_{m_3}({\bf k_3})
\qquad\quad (i=1,2)
\end{eqnarray}
where $I_{m_j}({\bf k_j})$ (j=2,3) are the spectator
contributions in the momentum space and have been
defined in (2.11). $I_{m_1}^{(i)}({\bf k_1-q})$ is
the struck quark contribution and is projected
out from (2.10) by using (2.12). Their expressions
are given in Appendix I. It turns out that $I_{m_1}^{(i)}
({\bf k_1-q})$, $I_{m_2}({\bf k_2})$ and
$I_{m_3}({\bf k_3})$ are independent of the quark spin
projections $m_i$. Hence the subscript `$m_1$' in
$\sum\limits_{{\alpha}_1,m_1}
b_{{\alpha}_1,m_1}({\bf 1;23})$ can be omitted.
Using the bag-type quark wave function (see (10) in
Appendix I) and completing the integrals over
$\prod\limits_{j={\bf 1}}^{\bf 3}d^3{\bf k}_j$, we obtain

\begin{equation}
$$W_1(x,Q^2)=N(1+{\eta}/2)
\sum\limits_{1\rightarrow 2,3}C({\bf 1;23})
\sum\limits_{\alpha}b_{\alpha}({\bf 1;23})
I_{\alpha}(R_1;{\xi}_2,{\xi}_3)
$$
\end{equation}

\begin{equation}
$$W_2(x,Q^2)=[{\eta}(1+3{\eta}/2)/(1+{\eta}/2)]W_1(x,Q^2)
$$
\end{equation}
where $N$ is a normalization constant, ${\eta}\equiv
1/(1+{\nu}^2/Q^2)$, ${\xi}_i\equiv R_i/R_1$ (i=2,3),
$C({\bf 1;23})\equiv MR_1{\xi}_2^3{\xi}_3^3$ and
$I_{\alpha}(R_1;{\xi}_2,{\xi}_3)$ is a dimensionless
integral which presents the collective contribution
from three valence quarks. The detail expression of
$I_{\alpha}(R_1;{\xi}_2,{\xi}_3)$ is given in (17) and
(18) in Appendix I.

If we neglect the symmetry
breaking effects coming from spin-dependent interactions,
then $R_2=R_3=R_1=R$, ${\xi}_2={\xi}_3={\xi}_1=1$ and
$I_{\alpha}$ does not depend on $\alpha$, i.e.
quark flavor, then (3.2) and (3.3) reduce into the SU(6)
symmetric result \cite{wang83}. We note that in the SU(6)
symmetric limit, one has $\sum\limits_{\alpha}b_{\alpha}=1$
for the proton and $\sum\limits_{\alpha}b_{\alpha}=2/3$
for the neutron, it then leads to
$F_2^n(x)/F_2^p(x)$=2/3 \cite{jaffe75}.

It is easy to verify, from (3.2)and (3.3),
that in the Bjorken limit, ${\eta}\rightarrow
{2Mx}/{\nu}\rightarrow 0$ and
$W_2/W_1={\eta}(1+3{\eta}/2)/(1+{\eta}/2)\rightarrow
{2Mx}/{\nu}$. Hence one obtains
$${\nu}W_2\rightarrow 2xMW_1\qquad i.e.\quad F_2(x)=2xF_1(x)$$
this implies that in the Bjorken limit the Callen-Gross relation
is satisfied within the model. Furthermore, in Appendix I we
briefly demonstrate the scaling behaviour of the structure
functions, {\it i.e.} both $F_1(x)$ and $F_2(x)$ are scaling
in the Bjorken limit and vanish when $x\rightarrow 1$. Numerical
calculations also confirm this conclusion.

Since $b_{\alpha}$ in (3.2) is the matrix element of
${\hat e}_q^2$, it is easy to rewrite (3.2) into

\begin{equation}
$$F_1(x)={1\over 2}\sum\limits_qe_q^2f_q(x)
$$
\end{equation}
where the quark distribution function $f_q(x)$ is
determined by the model result $I_{\alpha}(R_1;{\xi}_2,{\xi}_3)$.
Similarly we have

\begin{equation}
$$F_2(x)=x\sum\limits_qe_q^2f_q(x)
$$
\end{equation}
these are the parton model formulae as expected,
because the impulse approximation ($\bf a$) used in
our model is the same approximation used in the
parton model. In fact, the parton model formulae
(3.4) and (3.5) can also be obtained in any nucleon
model as far as the impulse approximation is
imposed. However, the magnitude and shape of
the quark distributions depend on approximations
($\bf b$) and ($\bf c$).
We note that in our model only three valence quark
configuration is considered, hence the model result
of the quark distributions should be identified with
the valence part only.

\subsection{Valence and Sea Decomposition}

For later analysis, we divide
the quark distribution functions $q(x)$ into two
parts, one that comes from the three valence quark
configurations which the primitive quark model
requires, the other from contributions of additional
effects:

\begin{equation}
$$q_{i}(x)=q_{iv}(x)+q_{is}(x);\quad
{\bar q}_{i}(x)={\bar q}_{iv}(x)+{\bar q}_{is}(x)
$$
\end{equation}
where the subscripts `$v$' and `$s$' denote the
terms `$valence$' and `$sea$'. The `$sea$' contributions
will be evaluated by using the QCD evolution approach
with suitable inputs at some low $Q^2$ scale.
By definition, we have ${\bar q}_{iv}(x)=0$
(${\bar q}_i={\bar u},{\bar d},{\bar s}$), $s_v(x)=0$ and
$q_{is}(x)={\bar q}_{is}(x)$. Then we have

\begin{equation}
$$F_2^p(x)=[F_2^p(x)]_{val}+[F_2^p(x)]_{sea}
$$
\end{equation}
where

\begin{equation}
$$[F_2^p(x)]_{val}=x[{4\over 9}u_v^p(x)+{1\over 9}d_v^p(x)]
  $$
\end{equation}
and

\begin{equation}
$$[F_2^p(x)]_{sea}=2x[{4\over 9}u_s^p(x)+{1\over 9}d_s^p(x)
+{1\over 9}s_s^p(x)]
$$
\end{equation}
small contributions coming from the charm quarks and
other heavier quarks are neglected. The equality
$q_s(x)={\bar q}_s(x)$ fulfills
the condition: the `sea' is flavorless, {\it i.e.}
the number of sea-quarks with flavor $f$ must be equal
to the number of sea-antiquarks with the same flavor.
It should be noted that this condition only requires a
weaker equality $\int_0^1q_s(x)dx=\int_0^1{\bar q}_s(x)dx$.

Using symmetry breaking SU(6) wave function with three-quark
configuration, we obtain

\begin{equation}
$$u_v^p(x)=4NMR{\xi}^3I_u(R;1,{\xi})
$$
\end{equation}
and

\begin{equation}
$$d_v^p(x)=2NMR{\xi}^{-3}I_d({\xi}R;{\xi}^{-1},{\xi}^{-1})
$$
\end{equation}
where $R=R_u^p$. The integral $I_u(R;1,{\xi})$ denotes
$I_{\alpha}(R_u^p;{\xi}_2,{\xi}_3)$, in (3.2), with
${\alpha}=u$, $R_1=R_u^p$, ${\xi}_2=1$ and
${\xi}_3=R_d^p/R_u^p\equiv \xi$ ($\xi\leq 1$). The same
notation is used for $I_d({\xi}R;{\xi}^{-1},{\xi}^{-1})$.
Obviously, the $I_u$ comes from the quark configuration
$[{\bf u;u,d}]$ or $[{\bf u;d,u}]$ and $I_d$ comes from
the quark configuration $[{\bf d;u,u}]$. In obtaining
(3.10) and (3.11), the symmetry
$I_{u,d}({\bf 1;2,3})=I_{u,d}({\bf 1;3,2})$
has been used, which can be verified from the explicit
expression of $I_{\alpha}(R_1;{\xi}_2,{\xi}_3)$.
In a similar manner, we can obtain the valence quark
distributions $u^n_v(x)$ and $d^n_v(x)$ in the neutron
and the result shows that

\begin{equation}
$$u^p_v(x)=d^n_v(x)\ ;\qquad  {\rm and}
\qquad  d^p_v(x)=u^n_v(x)$$
\end{equation}
this is expected because the isospin symmetry for the
nucleon wave function is assumed. Hence, we will omit
the superscript $p$ and $n$ below and use $u_v(x)$
instead of $u^p_v(x)$ and so on. We note that in the SU(6)
symmetry limit, $\xi =1$, one obtains, from (3.10) and (3.11),
$u_v(x)=2d_v(x)$.

The calculation of the structure function or quark
distributions in the deep inelastic region is divided
into two steps: (i) The valence quark distribution
functions at $Q_0^2$=(0.9GeV/c)$^2$ are calculated
$nonperturbatively$ by using the nucleon model
(under approximations ${\bf (a)-(c)}$),
(ii) using QCD evolution approach to evolve
these (valence) quark distributions $perturbatively$ to
higher $Q^2$ region, $e.g.$ $Q^2\sim$4, 10.7 and 15
(GeV/c)$^2$, where the experiments were performed.

\subsection{Valence Quark Distributions}

For the nonperturbative calculation in the step (i),
we need to determine the parameters in the model. One
can see from (17) in Appendix I that there
are only $three$ parameters: $R$ ($=R_u^p$), $\xi$
($=R_d^p/R_u^p$) and $\epsilon$ ($=\mid {\bf p}\mid_{max}R$,
where $\mid {\bf p}\mid_{max}$ is the maximum value
of three momentum of the struck quark inside
the nucleon).
The former two have been determined from the fit of the
rms radius of the neutron and proton (cf eq.(8) in
\cite{song92})

\begin{equation}
$$<r_n^2>=-2[(1-{\xi}^2)/(4-{\xi}^2)][<r_p^2>-3/(2M^2)] $$
\end{equation}
and the ratio (cf eqs. (9) and (10) in \cite{song92})

\begin{equation}
$${{\mu}_n}/{{\mu}_p}=-(2/3)\{ 1-8MR(1-{\xi})/[4MR(8+{\xi})+c]\}
$$
\end{equation}
where $c$ is a constant. The results were $R=$5 (GeV/c)$^{-1}$
and $\xi=0.85$, in this paper we take the same values
in order to maintain the consistency. The third parameter
$\epsilon$ is a new one and depends on the maximum
momentum of the struck quark inside the nucleon. In the
zero-temperature Fermi gas model, $\mid {\bf p}\mid_{max}$
can be identified as the Fermi momentum, {\it e.g.} in Ref.
\cite{bickerstaff90}, $k_F=0.2228$ GeV/c
is chosen for $u-$quark and $k_F=0.1539$ GeV/c is that
for $d-$quark. In our model, we choose $\epsilon=3$
(for $R=$5 (GeV/c)$^{-1}$, it implies
$\mid {\bf p}\mid_{max}$=0.6 GeV/c to constrain
the valence quark distributions to satisfy the
normalization conditions:

\begin{equation}
$$\int_0^1u_v(x)dx=2   \qquad {\rm and}\qquad
\int_0^1d_v(x)dx=1
$$
\end{equation}
which are fulfilled within numerical error (see Table 1).

For the perturbative calculation, the QCD scale
parameter $\Lambda$ is taken to be 0.3 GeV/c. The
numerical results of $xu_v(x)$ and $xd_v(x)$ at $Q^2$=0.81
and 15 (GeV/c)$^2$ are shown in Fig. 1a,b. For comparison,
the results given in the Fermi gas model \cite{bickerstaff90}
and the constituent quark model \cite{traini92} are shown.
The main reason for choosing $Q_0^2$=0.81(GeV/c)$^2$
as the renormalization scale is that the
perturbative QCD can be used in a less ambiguous
way above this scale. The running coupling constant
$\alpha_s(Q^2)$ in the leading order in the
modified minimal subtraction (${\bar M}{\bar S}$)
scheme (all quarks are assumed to be massless) is

\begin{equation}
$$\alpha_s(Q^2)={4\pi}/[(11-2f/3)ln(Q^2/{\Lambda}^2)]
$$
\end{equation}
this leads to $\alpha_s\simeq 0.4-0.6$ for $f=3$
and the QCD scale parameter: $\Lambda=0.2-0.3$
GeV/c. We have tried other two different values
of $Q_0^2$: (0.8GeV/c)$^2$ and (1.0GeV/c)$^2$,
the result shows that the predictions
for $F_2^p$ and $F_2^n$ etc. are not too sensitive
to the $Q_0^2$ value provided that the normalization
conditions (3.15) are imposed.

\subsection{Results with Sea Contributions}

For the sea quark distributions, we may assume,
as many authors
\cite{meyer91,schreiber91,bate92,bickerstaff90,traini92}
do, that the sea and
gluon distributions are zero at some very low
$Q_0^2$, then use leading order QCD evolution
approach to obtain the sea distributions in the
higher $Q^2$ region. But it will face the ambiguity
of using perturbative QCD. On the other hand, if we
start from $Q_0^2\simeq$1.0(GeV/c)$^2$, then some
initial input distributions for sea quarks and
gluons at this scale are needed, but we do not
have any information about these inputs at this
scale. Hence we first evolve the valence quark
distributions to $Q^2=4$ (GeV/c)$^2$, then
combine with the usual parameterized inputs of
gluon and sea quark distributions at this scale,
and obtain the sea distribution in the higher $Q^2$
region. Hence our sea quark and gluon distributions
are different from those in \cite{bickerstaff90}
and \cite{traini92}. However, taking a very low $Q_0^2$
(<0.1(GeV/c)$^2$) and assuming the initial sea
and gluon distributions are approximately zero at
this scale, we obtain, after evolution, almost the
same results as those given by \cite{bickerstaff90}
and \cite{traini92}.

The unpolarized structure functions $[F_2^p(x)]_v$,
$[F_2^p(x)]_{v+sea}$ and $[F_2^n(x)]_v$, $[F_2^n(x)]_{v+sea}$
are shown in Fig. 2a,b. The unpolarized sea and gluon
distributions given by this paper and other models are
shown in Fig. 3a,b, where $xq_s(x)=2x[u_s(x)+d_s(x)+s_s(x)]$
is singlet quark distribution. To separate the
different flavor contributions in the sea, we choose two
options:

\begin{equation}
$$({\bf i})\quad u_s(x)=d_s(x),\qquad {\rm or}\qquad
d_s(x)-u_s(x)=0
$$
\end{equation}
\begin{equation}
({\bf ii})\quad u_s(x)<d_s(x),\qquad {\rm or}\qquad
d_s(x)-u_s(x)>0
$$
\end{equation}
both options require $s_s(x)=0.25[u_s(x)+d_s(x)]$.
Option ({\bf i}) implies that sea violates SU(3) flavor
symmetry but maintains SU(2) symmetry, while option
({\bf ii}) means the sea even violates SU(2) flavor
symmetry. The numerical results for the integrals
$\int_0^1xu_vdx$, $\int_0^1xd_vdx$ and $\int_0^1xq_sdx$
are listed in Table 1. The
comparisons of our $F_2^p$ and $F_2^n$ with other models
are shown in Fig. 4a,b, where an SU(2) flavor asymmetric
sea, option ($\bf ii$), has been used in our prediction.
The result of $F_2^p(x)-F_2^n(x)$ and $[F_2^p(x)-F_2^n(x)]/x$
are shown in Fig. 5a and Fig. 5b. A comparison of
$F_2^p(x)-F_2^n(x)$ and $F_2^n(x)/F_2^p(x)$
given by different models are shown in Fig. 6a and Fig. 6b
respectively.

\subsection{Several Remarks}

{}From the results given above, we would like to make some
remarks:

(i) One can see from Fig. 1a,b that qualitatively the
results given by different models are the same. For $xu_v$,
our result gives a better fit to the data than the Fermi
gas model \cite{bickerstaff90} and constituent quark model
\cite{traini92}. For $xd_v$, however, their results are
better. As a consequence, our $F_2^p$ agrees well with
the data (Fig. 4a,b), but the $F_2^n$ is somewhat higher
than data in the region $x>0.3$ (Fig. 4b). For $0.1<x<0.3$,
both $F_2^p$ and $F_2^n$ given by our model seem to be
better than those from other two models, this is
due to our sea contribution is closer to the data
(Fig. 3a) in this $x$ range. For very low $x$ ($x<0.1$),
the model dependence is clearly seen, however due to
large theoretical uncertainties and data errors one
can not make a meaningful comparison between model
results and data.

(ii) In the parton model, the integral
${\rm I}_{\rm Gott.}\equiv\int_0^1dx[F_2^p(x)-F_2^n(x)]/x$
would be 1/3 if the sea is SU(2) flavor symmetric. However,
the NMC data shows ${\rm I}_{\rm Gott.}$=0.240$\pm$0.016.
One can see from Fig. 5b that the model result of the
valence quark contribution $[F_2^p(x)-F_2^n(x)]_{val}/x$
(dashed line) is higher than the data in the low-$x$ region,
while the sea contribution (dot-dashed line) given by (3.18),
{\it i.e.} option ($\bf ii$) leads to $[F_2^p(x)-F_2^n(x)]_{sea}/x
<0$ and the sum of valence and sea contributions is
consistent with the NMC data (solid line in Fig. 5b or Fig. 5a).
The numerical results for the integrals are listed in
Table 1. Within our model, the difference between
$F_2^{p,n}$ given by using symmetric sea (option $\bf i$)
and asymmetric sea (option $\bf ii$) is quite small
and only appears in the small-$x$ region ($x<0.2$).
However, this small difference could be the source which
causes the violation of the Gottfried sum rule.

(iii) The NMC data seems to favor a flavor asymmetric
sea. The idea that the sea might not be SU(2) symmetric
was suggested by Feynman and Field \cite{feynman77}
based on the Pauli exclusion principle. Since the proton
contains two valence up-quarks but only one down-quark, the
Pauli principle would suppress the creation of $u\bar u$
pairs relative to $d\bar d$ pairs. Ross and Sachrajda
\cite{ross79} showed that a nonzero value for $u_s(x)-d_s(x)$
can be obtained from higher order QCD contributions.
However, the effect given by perturbative QCD calculations
is too small to explain the deviation of Gottfried
sum rule (some earlier works \cite{donoghue77,he84}
also found that the up-sea quark component is not
identical with down-sea component within the proton,
the up-sea is larger than down-sea). This means that
a significant contribution should come from non-perturbative
interactions. Many different explanations for the
violation of Gottfried sum rule, have been suggested
recently, an incomplete list includes Ref.
\cite{sullivan72,henley90,stern91,signal91,kumano91,anselmino92}.
We note that, however, the
experimental data have large errors in the small-$x$
region, where the sea is the dominate contribution,
hence more precise measurements of $F_2^p(x)$ and
$F_2^n(x)$ at small-$x$ region are needed in order
to verify if and to what extent the Gottfried sum
rule is violated.

(iv) The ratio $F_2^n(x)/F_2^p(x)$, as shown in Fig. 6b,
is sensitive to the spin dependent interactions.
As discussed in \cite{song92} and \cite{sw92}
that the repulsion between $u-u$ pair comes from color
magnetism which is based on perturbative QCD and derived
from one-gluon exchange Breit-Fermi interactions, while
the attraction between $u-d$ quark pair can only be
induced by non-perturbative mechanism, {\it e.g.} instanton
interactions \cite{schuryak89}. One can see from
Fig. 6b that for $\xi=0.85<1$ the model gives a reasonable
$x$-behaviour of $F_2^n(x)/F_2^p(x)$ (except large-$x$
region), while for $\xi=1.15$ (which simulates a larger
spatial size for the scalar $u-d$ quark pair than that
for the vector quark pairs, that means the interaction
between $u-d$ quark pair is more repulsive than that
between $u-u$ or $d-d$ quark pairs) the prediction
seems to conflict with the data. Similar result is also
obtained in the light-cone constituent quark model (see
\cite{weber93}). It seems to imply
that to obtain a correct large $x$-behaviour of
$F_2^n(x)/F_2^p(x)$, in addition to the one-gluon
exchange effect other mechanism or non-perturbative
contributions are needed.

\section{SPIN-DEPENDENT STRUCTURE FUNCTIONS.}

\subsection{Formalism}

The spin-dependent structure functions $g_1$ and $g_2$
can be extracted from the antisymmetric part of the
hadronic tensor $W_{\mu\nu}^{(A)}$, eq.(2.4). For
convenience, we define two dimensionless quantities:
${\pi}^{\mu}\equiv P^{\mu}/M$ and ${\rho}^{\mu}\equiv
q^{\mu}/{\nu}$, then we have ${\pi}^2=1$, ${\rho}^2=
-Q^2/{\nu}^2$, $S\cdot {\pi}=0$ and ${\rho}\cdot {\pi}=1$.
Using these variables, (2.4) can be rewritten as

\begin{equation}
$$W_{\mu\nu}^{(A)}={\epsilon}_{\mu\nu\lambda\sigma}
 {\rho}^{\lambda}\{ S^{\sigma}g_1(x, Q^2)+[S^{\sigma}
-({{\rho}\cdot S}){\pi}^{\sigma}]g_2(x, Q^2)\}
$$
\end{equation}
On the other hand, the hadronic tensor can be
calculated from the dynamical model of the nucleon
in which the hadronic electromagnetic current and
the nucleon state are known. From (2.9) and (2.12),
we can write
\begin{equation}
$$W_{\mu\nu}^{(A)}(P,q,S)={\epsilon}_{\mu\nu\lambda\sigma}
{\rho}^{\lambda}I^{\sigma}(x, Q^2)
$$
\end{equation}
where $I^{\sigma}(x, Q^2)$ is obtained from (2.9)
by changing $I_{m_1}({\bf k_1-q})$ into
$I_{m_1}^{(A){\sigma}}({\bf k_1-q})$ and

\begin{equation}
$$I_{m_1}^{(A){\sigma}}({\bf k_1-q})=
(2{\pi})^3{\bar {\phi}}_{m}
({\bf k_1-q}){\gamma}^{\sigma}{\gamma}^5{\phi}_{m}
({\bf k_1-q})
$$
\end{equation}
Comparing (4.1) and (4.2), $g_1$ and $g_2$ can be
easily obtained.

In the nucleon rest frame: ${\pi}^{\mu}=(1,0,0,0)$,
$S^{\mu}=(0,0,0,1)$ and ${\rho}^{\mu}=(1,{\bf q}/{\nu})$.
Here the nucleon spin is chosen in the $z$-direction and
$\mid {\bf q}\mid$=${\nu}(1+{Q^2}/{\nu}^2)^{1/2}$.
For the direction of ${\bf q}$, we can choose:

\begin{eqnarray}
({\bf i})\ {\bf q}&\parallel& {-\bf S}\quad
{\rm nucleon}\ {\rm longitudinally}\ {\rm polarized}\
{\rm along}\ {\bf q}\nonumber \\
({\bf ii})\ {\bf q}&\perp& {\bf S}\qquad
{\rm nucleon}\ {\rm transversely}\ {\rm polarized}\
{\rm along}\ {\bf q}\nonumber
\end{eqnarray}

In the Bjorken limit, one has $({\bf i})$
${\rho\cdot S}=1$ for ${\bf q}\parallel {-\bf S}$,
$({\bf ii})$ ${\rho\cdot S}=0$ for ${\bf q}\perp {\bf S}$.
{}From these we have

\begin{equation}
$$g_1=I_L^0+I_L^3\equiv g_L
$$
\end{equation}
({\it i.e.} light-cone ``+'' component of $I_L$) and

\begin{equation}
$$g_1+g_2=I_T^3\equiv g_T
$$
\end{equation}
where $I_L^{0,3}$ denote $I^{\sigma}$ (${\sigma}=0,3$)
in which the $\bf q$ is chosen in the opposite
$z$-direction and $I_T^3$ denotes $I^{3}$ in which
the $\bf q$ is chosen in the opposite $x$-direction.
It is obvious that $g_1$ or $g_L$ depends only on the
``$longitudinal$'' scattering and $g_1+g_2$ or $g_T$
depends only on the ``$transverse$'' scattering,
while $g_2$ is the difference between ``$transverse$''
scattering and ``$longitudinal$'' scattering. In the
SU(6) symmetric limit, no special direction can be
assigned for the nucleon system, hence the rotational
invariance requires that the ``$transverse$''
scattering and ``$longitudinal$'' scattering are equal
and the difference between them, which is $g_2$, would
be zero \cite{feynman72}. However, the symmetry breaking
effects lead to a nonvanishing $g_2$.

It can be shown from (2.9) and (2.12) that

\begin{equation}
$$W_{\mu\nu}^{(A)}(P,q,S)={\epsilon}_{\mu\nu\lambda\sigma}
{\rho}^{\lambda}\{ 2(N/{\pi})C_0{\nu}
\sum\limits_{1\rightarrow 2,3}
\sum\limits_{\alpha_1 ,m_1}b_{\alpha_1 ,m_1}({\bf 1;23})
C({\bf 1;23})I_{\alpha_1 ,m_1}^{\sigma}({\bf 1;23})\}
$$
\end{equation}
where $I_{\alpha ,m}^{\sigma}({\bf 1;23})$ (${\sigma}$=0,3)
are the dimensionless integrals and given in Appendix II.
Since $I_{\alpha ,m}^{\sigma}({\bf 1;23})$ depend on sgn($m$),
{\it i.e.} the sign of spin projection $m$ of the
struck quark, it naturally leads to the difference between
spin-up and spin-down components, {\it i.e.} the spin-dependent
structure functions. Comparing (4.6) with (4.2),
one can obtain $I_L^{0,3}$ and $I_T^{3}$, then $g_1$ and $g_2$.
Their expressions are listed in Appendix II.

Similar to the unpolarized case, from the
term $\sum\limits_{{\alpha}_1 ,m_1}b_{{\alpha}_1 ,m_1}({\bf 1;23})
{\rm sgn}(m_1)$ in (4.6) (see (24) and (28) in Appendix II)
one can rewrite $g_1(x)$ and $g_1(x)+g_2(x)$ into
\begin{equation}
$$g_1(x)={1\over 2}\sum\limits_{q}e_q^2
[q_{\uparrow\parallel}(x)-q_{\downarrow\parallel}(x)]
$$
\end{equation}
\begin{equation}
$$g_1(x)+g_2(x)={1\over 2}\sum\limits_{q}e_q^2
[q_{\uparrow\perp}(x)-q_{\downarrow\perp}(x)]
$$
\end{equation}
which determine the longitudinal and transverse spin
structure functions. Similar to the unpolarized case,
the scaling behaviour $g_1$ and $g_2$ can be obtained
in the Bjorken limit and they vanish when $x\rightarrow 1$.

\subsection{Valence Polarizations}

The results of $\Delta u_v(x,Q^2)$ and
$\Delta d_v(x,Q^2)$ at different $Q^2$ values
are shown in Fig. 7a,b. We emphasize that all
parameters used here are the same as those in the
unpolarized case: $R$=5 (GeV/c)$^2$, $\xi=0.85$
and $\epsilon=3$, even the normalization factor
N is unchanged. The valence components
$[xg_1^p(x)]_{val}$ and $[xg_1^n(x)]_{val}$ are
shown in Fig. 8a,b. One can see from Fig. 8a that
the valence contribution $[xg_1^p(x)]_{val}$ at
$Q^2=10.7$ (GeV/c)$^2$ (the dotted curve) is
consistent with the data in the range of
$x>0.3$. This implies that the model gives a
good description of the valence components
of the spin-dependent structure functions, the
difference between the $[xg_1^p(x)]_{val}$ and
data in the small-$x$ region ($x<0.3$) is
naturally attributed to the polarized sea
contributions. For comparison, the results given
by Ref. \cite{schreiber91} are shown in Fig. 8a,b
(dashed curve). One can see that the results are
very similar. For the $xg_1^p(x)$, both curves
are very close and consistent with data at $x>0.3$.
Although their $xg_1^n(x)$ is larger than ours
and becomes negative at $x<0.02$, both curves
are consistent with data within large errors.

\subsection{Polarized Sea and Flavor Decomposition of
the Proton Spin}

For the polarized sea quarks, no information about their
distribution functions can be used as inputs at
$Q_0^2\simeq $1 (GeV/c)$^2$ to evolve them to required
higher $Q^2$ region. On the other hand, if we start from
very low momentum transfer scale, {\it e.g.} $Q_0^2\leq
$0.1 (GeV/c)$^2$ and assume that the sea and gluon
polarizations are approximately zero at this scale, the
leading order QCD evolution would generate a positive
polarized sea at higher $Q^2$ scale which seems to be
inconsistent with the SMC result. Hence we resort to a
different analysis in this section.

Similar to the unpolarized case, from the quark-parton
model we can write

\begin{equation}
$$g_1^p(x)=[g_1^p(x)]_{val}+[g_1^p(x)]_{sea}
$$
\end{equation}
where

\begin{equation}
$$[g_1^p(x)]_{val}={1\over 2}[{4\over 9}c_u{\Delta u}_v(x)
+{1\over 9}c_d{\Delta d}_v(x)]
$$
\end{equation}
and

\begin{equation}
$$[g_1^p(x)]_{sea}=[{4\over 9}c_u{\Delta u}_s(x)+
{1\over 9}c_d{\Delta d}_s(x)+{1\over 9}c_s{\Delta s}_s(x)]
$$
\end{equation}
where $\Delta u_{v}(x)=u_{v\uparrow}(x)-u_{v\downarrow}(x)$,
$\Delta d_{v}(x)=d_{v\uparrow}(x)-d_{v\downarrow}(x)$
and $\Delta q_{s}(x)=\Delta {\bar q}_{s}(x)$ ($q=u, d, s$)
are assumed. In (6.2) and (6.3), the QCD radiative correction
factors $c_u=1-(c_f-1){\alpha}_s/({2\pi})$ and
$c_d=c_s=1-(2c_f-1){\alpha}_s/{\pi}$ are manifestly included,
where $c_f=(33-8f)/(33-2f)$ with $f-$the number of quark
flavors. Similar expressions for
$[g_1^n(x)]_{val}$ and $[g_1^n(x)]_{sea}$ can be obtained
by exchange $u\leftrightarrow d$.

{}From the model result we obtain

\begin{equation}
$$\Delta u_v\equiv \int_0^1 {\Delta u}_{v}(x)dx=1.0003,
\quad \Delta d_v\equiv \int_0^1 {\Delta d}_{v}(x)dx=-0.2217
$$
\end{equation}
we note that these values will be slightly modified if
different $Q_0^2$ are used. Hence we would like to put
$5\%$ theoretical uncertainties on (4.12) and only present
the uncertainties in the final result (see (4.18) below).
(4.12) gives

\begin{equation}
$$\int_0^1[g_1^p(x)]_{val}dx=0.1974 \quad
\int_0^1[g_1^n(x)]_{val}dx=0.0106
$$
\end{equation}
Hence

\begin{equation}
$$\int_0^1[g_1^p(x)-g_1^n(x)]_{val}dx=0.1868
$$
\end{equation}
It implies that the valence parts almost saturate the
Bjorken sum rule, {\it i.e.} the sea contributions to $g_1^p$
and $g_1^n$ are almost the same within the model. Using
the Bjorken sum rule
$\int_0^1[g_1^p(x)-g_1^n(x)]dx=1/6
({g_A}/{g_V})_{n\rightarrow p}(1-{\alpha}_s/{\pi}
-O({\alpha}_s^2))$=0.191 (where ${\alpha}_s\simeq$0.22
has been used) and the EMC data

\begin{equation}
$$\int_0^1g_1^p(x)dx=0.126\pm 0.018
$$
\end{equation}
one obtains

\begin{equation}
$$\int_0^1g_1^n(x)dx=-0.065\pm 0.018
$$
\end{equation}
Combining (4.13), (4.15) and (4.16) we have

\begin{equation}
$$\int_0^1[g_1^p(x)]_{sea}dx=-0.0714, \quad
\int_0^1[g_1^n(x)]_{sea}dx=-0.0756
$$
\end{equation}
Using the $\nu-p$ scattering data \cite{ahrens87}: $\Delta s\equiv
\Delta s_s+\Delta{\bar s}_s=-0.15\pm 0.09$, we obtain a
set of sea polarizations:

\begin{equation}
$$2\Delta u_s=-0.229\pm 0.021 \quad 2\Delta d_s=-0.257\pm 0.033
\quad 2\Delta s_s=-0.150\pm 0.090
$$
\end{equation}
{}From (4.18), we have $\Delta d_s\simeq 1.12\Delta u_s$,
$\Delta s_s\simeq 0.31(\Delta u_s+\Delta d_s)$ (these
asymmetric relations are very similar to those in the
unpolarized case). Assuming the polarized sea quarks
for different flavors have the same $x$-behaviour,
which can be parameterized as:

\begin{equation}
$$\Delta q_s(x)=a_q(2-x)(1-x)^{8.3} \quad (q=u,d,s)
$$
\end{equation}
where $a_u=-0.557$, $a_d=1.122a_u$ and $a_s=0.309(a_u+a_d)$.
Using (4.19), the $[xg_1^{p}(x)]_{sea}$ is calculated
and the result is shown in Fig. 9a (dot dashed curve),
where the dotted curve is valence component only.
The solid curve shows the sum of valence and sea
contributions. The corresponding curves for the
$xg_1^{n}(x)$ are shown in Fig. 9b. The result for
$g_1^n(x)$ is shown in Fig. 10a, data are taken from
[24,25]. The result for $xg_1^d(x)$ is shown in
Fig. 10b. It shows that the theoretical
predictions are consistent with data within large
errors. The numerical results and comparison with
other models are listed in Table 2.

\subsection{Several Remarks}

{}From the result given above, several observations are in order:

(i) In the static SU(6) limit, there are no sea contributions
and

\begin{equation}
$$\Delta u_v=4/3,\quad \Delta d_v=-1/3, \quad
\Delta u_v=-4\Delta d_v,\quad
\Delta u_v+\Delta d_v=1                           $$
\end{equation}
i.e. $100\%$ proton spin is contributed by the spin of the
(valence) quarks. However, due to the presence of the
quark-gluon interaction, which is described by the
quantum chromodynamics (QCD), quark-antiquark pairs
and gluons ($sea$) are produced with appropriate orbital
angular momentum in addition to the usual three valence
quarks in the $S-$state. It is obviously that the $valence$
quarks can no longer be responsible for carrying $100\%$
of the proton's spin. The result (4.12) shows

\begin{equation}
$$\Delta u_v\simeq 1.00,\quad\Delta d_v\simeq -0.22,\quad
\Delta u_v\simeq -4.5\Delta d_v,\quad \Delta u_v+\Delta d_v\simeq 0.78$$
\end{equation}
It implies that even in the high $Q^2$ (deep inelastic)
region the valence quarks' spin still contributes a
large fraction of the proton spin. This result agrees
with the earlier analysis by Close \cite{close90} (in
\cite{close90}
using $(g_A/g_V)_{n\rightarrow p}=1.257=
\Delta u_v-\Delta d_v$ and assuming $\Delta u_v=-4\Delta d_v$,
the author obtained $\Delta u_v\simeq 1.00$,
$\Delta d_v\simeq -0.25$ and $\Delta u_v+\Delta d_v\simeq 0.75$).

(ii) The net sea quark spin polarization is $-0.64$,
i.e $64\%$ of proton's spin comes from sea quarks and
the sea is polarized against the proton spin.
This means that although the valence quarks still
highly polarized, the sea quarks are spinning
in the opposite direction and cancel most of valence
polarization (the cancellation occurs at small-$x$ region,
see Fig. 9a). From (4.12) and (4.18), one obtains

\begin{equation}
$$\Delta u=0.771\pm 0.054,\quad \Delta d=-0.479\pm 0.035,\quad
\Delta s=-0.150\pm 0.090
$$
\end{equation}
and

\begin{equation}
$$\Delta u+\Delta d+\Delta s=-0.142\pm 0.110
$$
\end{equation}
So that only $14\%$ of the proton spin comes
from the spin of quarks and antiquarks. The rest $86\%$
must comes from other sources, {\it e.g.} the gluon polarization
and the orbital angular momentum of the quarks and gluons.

(iii) The angular momentum sum rule \cite{ratcliffe87} is

\begin{equation}
$${1\over 2}={1\over 2}\sum\limits_q {\Delta q}+
{\Delta g}+L_{q+g}                                   $$
\end{equation}
where the first term on the r.h.s. is the quark's
spin contribution, ${\Delta g}$ is the gluon spin
polarization and $L_{q+g}$ is the total orbital
angular momentum of the quarks and gluons. Since
$\sum\limits_q {\Delta q}\simeq 0.14$, we have
${\Delta g}+L_{q+g}\simeq 0.43$. If we assume that
${\Delta g}\simeq L_{q+g}\simeq 0.22$, then the
`$gluonic$' contribution \cite{altarelli88,efremov88}
(due to the Adler-Bell-Jackiw axial anomaly) to the
quark's spin polarization $\Delta\Gamma\equiv
-[\alpha_s(Q^2)/2\pi]\Delta g(Q^2)$ would be very small
($\simeq -0.009$). Hence our conclusion given above
will not be affected by adding this gluonic term,
except for the situation that $\Delta g(Q^2)$ is very large
({\it e.g.} see \cite{altarelli89}). It should be
noted that from leading order perturbative QCD
evolution ({\it e.g.} see \cite{song89}), as $Q^2$ grows,
the quark spin is conserved ($\Delta q$ does not vary)
and the gluon spin grows like $lnQ^2$ ($\Delta\Gamma$
does not vary as $\alpha_s(Q^2)\sim 1/lnQ^2$), the total
angular momentum conservation (4.24) requires $L_{q+g}$
must grow in the negative direction to cancel $\Delta g$.

(iv) It is easy to check that our result of flavor
decomposition of the proton spin (4.22) can also
accommodate the existing hyperon $\beta-$decay constants.
Using F+D=$\Delta u-\Delta d$, F$-$D=$\Delta d-\Delta s$
and (4.22) we have F=0.461, D=0.789 and F/D=0.584 (which
agrees well with data from \cite{hsueh88}). A comparison
of the predicted hyperon $\beta$-decay constants with data
and other models is shown in Table 2. The quark polarizations
(4.22) indeed gives a good fit to the existing data except for
$(g_A/g_V)_{\Xi^-\rightarrow \Lambda}$.

(v) Combining our model result (4.13) with Bjorken
sum rule and EMC data, the sea should be negatively
polarized as shown in (4.18). However, as we
mentioned above that a negatively polarized sea can
not be generated by leading order perturbative QCD
from very low $Q_0^2$ (if at this scale the sea and
gluon polarizations are zero) and it must come from
some nonperturbative mechanism. It is interesting to
note that a preliminary lattice QCD result reported
in \cite{dong93} shows that the sign of quark loop
(`sea') contribution to iso-scalar axial coupling
constant $g_A$ is $negative$. Another possible mechanism
which may cause a negatively polarized sea is
nonperturbative instanton interactions \cite{dorokhov90}.

(vi) Similar to the unpolarized case, the polarized sea
seems to be flavor asymmetric, see (4.18), at least
violates SU(3) flavor symmetry. In the unpolarized case,
it is relatively easy to verify if the unpolarized sea
violates SU(2) flavor symmetry by measuring the deviation
of the Gottfried sum rule, because the $u_v$ and $d_v$
must obey the valence quark number sum rules in (3.15).
In the polarized case no similar constraints
exist. Hence the valence part of the
$\int_0^1[g_1^p(x,Q^2)-g_1^n(x,Q^2)]dx$,
i.e. ${1\over 6}\int_0^1[\Delta u_v(x,Q^2)-\Delta d_v(x,Q^2)]dx$
is unknown. To determine the sign of the sea term
${1\over 3}\int_0^1[\Delta u_s(x,Q^2)-\Delta d_s(x,Q^2)]dx$,
one needs to measure $\Delta u_v(x,Q^2)-\Delta d_v(x,Q^2)$
independently in addition to $g_1^p$ and $g_1^n$.

\section{SECOND SPIN-DEPENDENT STRUCTURE FUNCTION.}

For the second spin-dependent structure function $g_2$,
we will not discuss the sea contributions and only
briefly present the results given by the valence
contribution, so the subscript `$val$' will be omitted
below. But bear in mind that these results, including
only valence quarks, may have substantial modifications
in the small-$x$ region as we have seen in the $g_1$ case.
The model results for $g_2^{(p,n)}$ are shown in Fig. 11a.
It can be shown that in the SU(6)
symmetric limit ($\xi$=1), $g_1=g_T$, hence
$g_2^{(p,n)}(x, Q^2)=0$ and $\int_0^1g_2^{(p,n)}(x, Q^2)dx=0$,
{\it i.e.} the Burkhardt-Cottingham (B-C) sum rule
is fulfilled. For SU(6) symmetry breaking wave function,
$\int_0^1g_2^p(x, Q^2)dx\simeq 0$,
{\it i.e.} the B-C sum rule is approximately fulfilled.
Similarly, $\int_0^1g_2^n(x, Q^2)dx$ is also very small
(10$^{-3}$) within the model. A comparison of $g_2^p(x)$
given by different models is shown in Fig. 11b. The
model dependence is easily seen, all models, however,
give qualitatively similar results, {\it i.e.} $g_2(x)$
starts positive from $x\simeq 0$, changes its sign at
some $x=x_0$ and after passing through a minimum it
tends to zero. There are two distinctions between our
result and those given by other models: (i) our $g_2$
is smaller than $(g_2)_{\rm MIT}$ and $(g_2)_{\rm SST}$
in the wide range of $x$ and (ii) our $x_0$ ($\sim$0.15)
is lower than $(x_0)_{\rm SST}$ ($\sim$0.25) and
$(x_0)_{\rm MIT}$ ($\sim$0.30).

Unlike the unpolarized structure functions
and $g_1$ of nucleon, the second spin structure
function $g_2(x)$ involves contributions from
the quark-gluon correlations and quark
mass effects even in the large $Q^2$ limit. In
the formalism of operator product expansion, these
contributions come from local operators having
twist-three. The structure function $g_2(x)$ can
be decomposed into the two different twist pieces:

\begin{equation}
$$g_2(x)=g_2^{WW}(x)+{\bar g}_2(x)
$$
\end{equation}
where \cite{wandzura77}

\begin{equation}
$$g_2^{WW}(x, Q^2)=-g_1(x, Q^2)+\int_x^1dyg_1(y, Q^2)/y
$$
\end{equation}
is twist-2 piece and

\begin{equation}
$${\bar g}_2(x, Q^2)=g_1(x, Q^2)+g_2(x, Q^2)
-\int_x^1dyg_1(y, Q^2)/y
$$
\end{equation}
is the twist-3 piece. The model results of $g_2^{WW}$
and ${\bar g}_2$ for the proton and neutron are shown
in Fig. 12a and Fig. 12b. Similar to other model
calculations, our result also shows that although
$g_2(x)$ is much smaller than $g_1(x)$, the twist-2
piece $g_2^{WW}(x)$ and twist-3 piece ${\bar g}_2(x)$
of the $g_2(x)$ are quite large, in particular in the
small-$x$ region. However, $g_2^{WW}$ and ${\bar g}_2$ a
lmost cancel and the sum of them gives a very small
$g_2$. So that the higher twist effects cannot be
neglected. Some numerical results are listed in
Table 3.

\section{SUMMARY}

In this paper both polarized and unpolarized structure
functions for a free nucleon are calculated at moderate
$Q^2$ value by using the modified CM bag model
and evolved to higher $Q^2$ region by using leading
order perturbative QCD. Despite some approximations
we have made, the model gives a reasonable
description for the valence quark distributions at
$x>0.3$. For the small $x$ region, the sea quark
distributions are needed. For the unpolarized case,
combining the valence and sea contributions, one can
accommodate the existing unpolarized data except
$F_2^n/F_2^p$ at the large $x$ region. The violation
of the Gottfried sum rule can be attributed to a flavor
asymmetric sea. For the spin-dependent structure
functions, the existing data can be also accommodated
within the model and a negatively polarized sea is
required.

For unpolarized sea, the
leading order QCD evolution leads to a good or at least
qualitative agreement with the data, if a moderate or
very low $Q_0^2$ are used. On the other hand, the
perturbative QCD is unable to give a negatively
polarized sea and various nonperturbative mechanisms
should be further studied. In addition, it seems very likely
that both unpolarized and polarized sea violate SU(2) flavor
symmetry. However, for the polarized sea, the uncertainties
of model result and data are too large to make a definite
conclusion.

The spin-dependent effect plays an important role
in accommodating the data. In the model calculation,
the magnitude and shape of $u_v$, $d_v$ and $\Delta u_v$,
$\Delta d_v$, in particular $F_2^n/F_2^p$, $g_1^n$ and
$g_2^{p,n}$ are sensitive to the symmetry breaking
effect which comes from spin-dependent interactions.
Without spin-dependent effect, we would obtain
$F_2^n/F_2^p$=2/3, $g_1^n$=0 and $g_2^{p,n}$=0 if
the sea contributions are ignored. Taking $\xi$=0.85
our result for $F_2^n/F_2^p$ can not accommodate the
data in the large-$x$ region.
Possible sources which cause this failure presumably
are (i) the cavity approximation for the quark wave
function is too sharp at the boundary and does not
have suitable `$soft$' behaviour, and (ii) The parameter
$\xi$ which simulates the symmetry breaking effects of
the nucleon wave function is not good enough to describe
different $x$-behaviour between $u_v$ and $d_v$.
Further studies on these problems are needed.

Finally, we note that only leading order perturbative
QCD evolution has been used. For the momentum transfer
scale larger than $Q_0^2$$\sim$($0.9-1$GeV/c)$^2$, we
expect that perturbative next-to-leading order QCD
corrections and higher twist effects do not significantly
change the leading order result. However, it is well known
that these corrections and higher order effects become more
and more important when $x$ approaches the end points 0 and
1. Hence our results are more reliable at the middle range
of the $x$ and less reliable at $x$ near the end points.

\acknowledgements

The authors acknowledge many useful conversations with
H. J. Weber. We also thank P. K. Kabir, R. Lourie, O.
Rondon-Aramayo and other members of the INPP for their
comments and suggestions. This work was supported by the
US Department of Energy and the Commonwealth Center for
Nuclear and High Energy Physics, Virginia, USA.

\vfill\eject

\appendix{Appendix I.}

We introduce the projection operators

\begin{equation}
$$\Lambda_{\mu\nu}^{(1)}=
[-g_{\mu\nu}+{\eta}P_{\mu}P_{\nu}/M^2]/2\ ;\quad
\Lambda_{\mu\nu}^{(2)}={\eta}[-g_{\mu\nu}+3{\eta}P_{\mu}P_{\nu}/M^2]/2
$$
\end{equation}
which satisfy
\begin{equation}
$${\Lambda^{(1)}}^{\mu\nu}W_{\mu\nu}=W_1\ ;\quad
{\Lambda^{(2)}}^{\mu\nu}W_{\mu\nu}=W_2
$$
\end{equation}
Using (2) and eq.(2.11) in section II, we obtain (3.1),
where
\begin{eqnarray}
I_{m_1}^{(1)}({\bf l})&=&
{\Lambda}^{(1)\mu\nu}I_{m_1,\mu\nu}({\bf l})=
(1-{\eta}/2)I_{m_1}({\bf l})
+{\eta}\cdot I_{m_1}'({\bf l})\\
I_{m_1}^{(2)}({\bf l})&=&
{\Lambda}^{(2)\mu\nu}I_{m_1,\mu\nu}({\bf l})
={\eta}[(1-{3\eta}/2)I_{m_1}({\bf l})
+3{\eta}\cdot I_{m_1}'({\bf l})]
\end{eqnarray}
where ${\eta}=1/(1+{\nu}^2/Q^2)$,
${\bf l}\equiv {\bf q-k_1}$ and

\begin{eqnarray}
I_{m_1}({\bf l})&=&(2{\pi})^3{\bar{\phi}}_{m_1}
({\bf l}){\rlap/k_1}{\phi}_{m_1}({\bf l})\\
I_{m_1}'({\bf l})&=&(2{\pi})^3
{\bar{\phi}}_{m_1}({\bf l}){\rlap/P}
{\phi}_{m_1}({\bf l})(k_1\cdot P/M^2)
\end{eqnarray}
In the nucleon rest frame, $I_{m_1}({\bf l})=I_{m_1}'({\bf l})$,
hence we have

\begin{eqnarray}
I_{m_1}^{(1)}({\bf l})&=&(1+{\eta}/2)
I_{m_1}({\bf l})\\
I_{m_1}^{(2)}({\bf l})&=&{\eta}(1+{3\eta}/2)
I_{m_1}({\bf l})
\end{eqnarray}

Until now we have not yet chosen the specific form of the
rest frame quark wave function $q_m({\bf r})$. To maintain
the consistency, we use the same quark wave function as
in [45], {\it i.e.} the cavity approximation

\begin{equation}
q_m({\bf r})=N({\omega})^{-1/2}\left(
\begin{array}{c}
ij_0({\omega}{r/R})U_m\\
-ij_1({\omega}{r/R}){\vec \sigma}\cdot {\hat {\bf r}}U_m
\end{array}\right)
\end{equation}
and

\begin{equation}
{\phi}_{m}({\bf k})=
(2/{\pi})^{1/2}{iR^3}[N({\omega})]^{-1/2}
\left(
\begin{array}{c}
t_0(\omega, kR)U_m\\
t_1(\omega, kR){\vec \sigma}\cdot {\hat {\bf k}}U_m
\end{array}\right)
\end{equation}
where $t_i({\alpha}, {\beta})=\int_0^1x^2dxj_i({\alpha}x)
j_i({\beta}x)$ (i=0,1), $\omega=$2.04,
${\hat {\bf k}}={\bf k}/|{\bf k}|$,
$R$ is the bag radius. The normalization factor is
$N({\omega})=4{\pi}{R}^3[1-j_0^2({\omega})]/{\omega}^2$.
Using (10), the spectator
quark pieces (2.11) can be rewritten as

\begin{equation}
$$I_{m_j}({\bf k_j})=C_j
2k_j[t_0({\omega},{k_jR_j})+t_1({\omega},{k_jR_j})]^2
\qquad (j=2,3)
$$
\end{equation}
and the struck quark pieces (7) and (8) become

\begin{eqnarray}
I_{m_1}^{(1)}({\bf k_1-q})&=&C_1(1+{\eta}/2)
2k_1[t_0^2({\omega},{lR_1})+t_1^2({\omega},{lR_1})]\\
I_{m_1}^{(2)}({\bf k_1-q})&=&C_1{\eta}
(1+3{\eta}/2)2k_1[t_0^2({\omega},{lR_1})
+t_1^2({\omega},{lR_1})]
\end{eqnarray}
where $C_i=8{\pi}^2{R_i^6}/N({\omega})$ (i=1,2,3).
\smallskip
For brevity, we define the dimensionless
variables:\ ${\beta}_i\equiv k_iR_1$ (i=1,2),\
${\beta}_0\equiv l_0R_1=\mid (M+q_0)R_1-{\beta}_1\mid$,\
${\beta}_3\equiv {\beta}_0-{\beta}_2$,
$\beta\equiv \mid{\bf q-k_1}\mid R_1$,
${\delta}\equiv \mid {\bf q}\mid R_1$ and
${\epsilon}=({\bf p}_1)_{max}R_1$ where
$({\bf p}_1)_{max}$ is the maximum value
of three momentum of the struck quark inside
the nucleon. Using these variables, we have

\begin{equation}
$$\int d^3{\bf k}_2d^3{\bf k}_3{\delta}^4(q+p-\sum\limits_i{\bf k}_i)
\rightarrow {2\pi}/({\beta}R^2)
\int\limits_{({\beta}_0-{\beta})/2}^{({\beta}_0+{\beta})/2}
d{\beta}_2\cdot {\beta}_2({\beta}_2-{\beta}_0)
$$
\end{equation}
and

\begin{equation}
$$\int d^3{\bf k}_1\rightarrow ({2\pi}/{\delta})
\int\limits_{\delta -\epsilon}^{\delta +\epsilon}
{\beta}_1d{\beta}_1\int\limits_0^{{\beta}_0}{\beta}d{\beta}
$$
\end{equation}

Substituting (11) and (12) into (3.1), using
(14) and (15) we finally obtain

\begin{equation}
$$W_1(x,Q^2)=N(1+{\eta}/2)
\sum\limits_{1\rightarrow 2,3}\sum\limits_{\alpha}
b_{\alpha}({\bf 1;23})C({\bf 1;23})
I_{\alpha}(R_1;{\xi}_2,{\xi}_3)
$$
\end{equation}
{\it i.e.} (3.2), where

\begin{equation}
$$I_{\alpha}(R_1;{\xi}_2,{\xi}_3)=
\int\limits_{\delta -\epsilon}^{\delta +\epsilon}
{\beta}_1d{\beta}_1
\int\limits_0^{{\beta}_0}({\beta}/{\delta})d{\beta}
[t_0^2({\omega}, {\beta})+t_1^2({\omega}, {\beta})]
I_{\alpha}^{({\bf 23})}({\beta}_0, {\beta})
$$
\end{equation}
and
\begin{equation}
$$I_{\alpha}^{({\bf 23})}({\beta}_0, {\beta})=
\int\limits_{({\beta}_0-{\beta})/2}^{({\beta}_0+{\beta})/2}
({\beta}_2/{\beta})({\beta}_2-{\beta}_0)d{\beta}_2
\prod\limits_{i=2}^3[t_0({\omega},{\xi}_i{\beta}_i)
+t_1({\omega},{\xi}_i{\beta}_i)]^2
$$
\end{equation}
where $t_i(a,b)=\int_0^1x^2dxj_i(ax)j_i(bx)$ (i=0,1)
and $\omega=$2.04. Similarly, we can obtain (3.3).

We note that although the quark wave function (9),
or (10), has been used, other options are allowed.
For instance, a Gussian-type wave function [70] (The
authors thank H. J. Weber for pointing out this option)

\begin{equation}
q_{m}({\bf r}_i)=({\alpha}/{\sqrt \pi})^{1/2}
e^{{\alpha}^2R^2/2}
\left(
\begin{array}{c}
U_m\\
i({\alpha}^2/2m_q){\vec \sigma}\cdot{\vec {\bf r}}_iU_m
\end{array}
\right) e^{-{\alpha}^2r_i^2/2}
\end{equation}
and its Fourier transformation
\begin{equation}
{\phi}_{m}({\bf k}_i)=
1/({\alpha}^2{\sqrt \pi})
e^{{\alpha}^2R^2/2}\left(
\begin{array}{c}
U_m\\
(1/2m_q){\vec \sigma}\cdot{\vec {\bf k}}_iU_m
\end{array}\right) e^{-{p_i^2}/2{\alpha}^2}
\end{equation}
can be used to calculate the structure functions.

In the Bjorken limit,
${\nu}/{\mid {\bf q}\mid}\rightarrow 1$,
$\eta\rightarrow 0$ and $\mid {\bf q}\mid\simeq q_0+Mx$,
then $\beta_0=(M+q_0-\mid{\bf q}\mid)R_1\rightarrow MR_1(1-x)$,
hence the integral
$\int_0^{{\beta}_0}{\beta}d{\beta}
[t_0^2({\omega}, {\beta})+t_1^2({\omega}, {\beta})]
I_{\alpha}^{({\bf 23})}({\beta}_0, {\beta})$ depends only on $x$.
Considering that the initial quark momentum is much smaller
than the virtual photon momentum, i.e.
$\mid {\bf p}_1\mid << \mid {\bf q}\mid$, therefore
$\epsilon << \delta$, one can
see that the ${\beta}_1$ value is restricted
in a very narrow region around $\delta$:
${\delta \pm \epsilon}$, and the integral over ${\beta}_1$
can be approximately given by the mean-value theorem.
Then (17) becomes

\begin{equation}
$$I_{\alpha}(R_1;{\xi}_2,{\xi}_3)\simeq
2{\epsilon}\cdot \int_0^{{\beta}_0}{\beta}d{\beta}
[t_0^2({\omega}, {\beta})+t_1^2({\omega}, {\beta})]
I_{\alpha}^{({\bf 23})}({\beta}_0, {\beta})
$$
\end{equation}
which depends only on the variable $x$, hence both
$F_1(x)$ and $F_2(x)$ are scaling in the Bjorken
limit and vanish when $x\rightarrow 1$. Numerical
calculations also
confirm this conclusion.

\appendix{Appendix II.}

{}From (2.9) and (2.12), we have

\begin{eqnarray}
W_{\mu\nu}^{(A)}(P,q,S)&=&{\epsilon}_{\mu\nu\lambda\sigma}
k_1^{\lambda}\sum\limits_{1\rightarrow 2,3}
\sum\limits_{\alpha_1,m_1}b_{\alpha_1,m_1}({\bf 1;23})
M(4{\pi}^2{R}_1)^{-3}
\int\prod\limits_{i=1}^3d^3{\bf k}_i/(2k_i)\nonumber \\
&\cdot& {\delta}^4(q+P-{\sum\limits_ik_i})
I_{m_1}^{(A)\sigma}({\bf k_1-q})
I_{m_2}({\bf k_2})I_{m_3}({\bf k}_3)
\end{eqnarray}
where the axial-vector current piece coming from
the struck quark $I_{m_1}^{(A)\sigma}({\bf k_1-q})$
has been given in (4.3).
Using (10), it is easy to show that

\begin{equation}
{\bar {\phi}}_{m}({\bf k_1-q}){\gamma}^{\sigma}
{\gamma}^5{\phi}_{m}({\bf k_1-q})
=(2/{\pi})(R_1^6/N){\rm sgn}(m)
\left\{
\begin{array}{c}
-2({\beta}_z/{\beta})t_0t_1\qquad\qquad
\sigma=0 \\
t_0^2+(2{\beta}_z^2/{\beta}^2-1)t_1^2
\qquad \sigma=3
\end{array}\right\}
\end{equation}
where ${\beta}_z=|({\bf q-k_1})_z|R_1$.

Substituting (23) and the
spectator quark pieces (11) into (22) and using
(14) and (15), we arrive at (4.6). The results for
$I_L^{0,3}$, $I_T^{3}$ and $g_1$, $g_2$ are listed below:

(i) For ${\bf q}\parallel {-\bf S}$:

\begin{equation}
$$I_L^{i}=N{\pi}\sum\limits_{{\bf 1}\rightarrow {\bf 2,3}}
\sum\limits_{{\alpha}_1 ,m_1}b_{{\alpha}_1 ,m_1}({\bf 1;23})
C({\bf 1;23})
{\rm sgn}(m_1)I_{L}^{(i)}({\beta}_0,\delta ,\epsilon)\quad
(i=0,3)
$$
\end{equation}
with
\begin{equation}
$$I_{L}^{(0)}({\beta}_0,\delta ,\epsilon)=
\int_{\delta -\epsilon}^{\delta +\epsilon}d{\beta}_1
\int_0^{{\beta}_0}d{\beta}(-2{\beta}_z)
t_0({\omega}, {\beta})t_1({\omega}, {\beta})
I_{\alpha_1}^{({\bf 23})}({\beta}_0,{\beta})
$$
\end{equation}
and
\begin{equation}
$$I_L^{(3)}({\beta}_0,\delta ,\epsilon)=
\int_{\delta -\epsilon}^{\delta +\epsilon}d{\beta}_1
\int_0^{{\beta}_0}{\beta}d{\beta}
[t_0^2({\omega}, {\beta})+(2{\beta}_z^2/{\beta}^2-1)
t_1^2({\omega}, {\beta})]I_{\alpha_1}^{({\bf 23})}({\beta}_0, {\beta})
$$
\end{equation}
where ${\beta}_z=({\beta}_1^2-{\delta}^2-{\beta}^2)/{2\delta}$
and the spectator quark contribution
$I_{\alpha}^{({\bf 23})}({\beta}_0,{\beta})$ is the same as that
in the unpolarized case, i.e. (18).

It can be shown

\begin{equation}
$$g_1=(1-{\eta})[I_L^3+I_L^0/{\sqrt{1-\eta}}+
I_T^3{\eta}/(1-{\eta})]
$$
\end{equation}
In the Bjorken limit, (27) reduces into (4.4).

(ii) For ${\bf q}\perp {\bf S}$:

\begin{equation}
$$I_T^3=N{\pi}\sum\limits_{{\bf 1}\rightarrow {\bf 2,3}}
\sum\limits_{{\alpha}_1 ,m_1}b_{{\alpha}_1 ,m_1}({\bf 1;23})
C({\bf 1;23}){\rm sgn}(m_1)I_T^{(3)}({\beta}_0,\delta ,\epsilon)
$$
\end{equation}
where

\begin{equation}
$$I_T^{(3)}({\beta}_0,\delta ,\epsilon)=
\int_{\delta -\epsilon}^{\delta +\epsilon}d{\beta}_1
\int_0^{{\beta}_0}d{\beta}
\{ t_0^2({\omega}, {\beta})
+[({\beta}_1^2-{\beta}_{1x}^2)/{\beta}^2-1]
t_1^2({\omega}, {\beta})\} I_{\alpha}^{({\bf 23})}({\beta}_0, {\beta})
$$
\end{equation}
where ${\beta}_{1x}=({\beta}^2-{\delta}^2-{\beta}_1^2)/{2\delta}$.
Similarly, we have

\begin{equation}
$$g_2=(1-{\eta})[I_T^3-I_L^3-I_L^0/{\sqrt{1-\eta}}]
$$
\end{equation}
which, combining (27), reduces into (4.5) in the Bjorken limit.

\figure{Fig. 1a. The calculated $xu_v(x,Q^2)$ at
$Q^2$=0.81 (GeV/c)$^2$ (solid line) and QCD evolved
result at $Q^2$=15 (GeV/c)$^2$ (dashed line) given
in this paper. Data is taken from Ref. [1], The
results given by Fermi gas model [42] (dotted
line) and constituent quark model [43] (dot-dashed
line) are also shown.}

\figure{Fig. 1b. Same as Fig. 1a, but for
$xd_v(x,Q^2)$.}

\figure{Fig. 2a. The calculated $[F_2^p(x, Q^2)]_{val}$
at $Q^2$=0.81 (GeV/c)$^2$ (solid line) and QCD evolved
result at $Q^2$=15 (GeV/c)$^2$ (dashed line).
$F_2^p(x, Q^2)$ (dotted line: valence + symmetric sea;
dot-dashed line: valence + asymmetric sea) are compared
with data [3].}

\figure{Fig. 2b. Same as Fig. 2a, but for $F_2^n$.}

\figure{Fig. 3a. The calculated $xq_s(x)$ given in
this paper (solid line) compared with data [1] at
$Q^2$=15 (GeV/c)$^2$. The results given by Fermi
gas model [42] (dotted line) and constituent quark
model [43] (dot dashed line) are also shown.}

\figure{Fig. 3b. Same as Fig. 3a, but
for the unpolarized gluon distribution $xG(x)$.}

\figure{Fig. 4a. Comparison of our $F_2^p(x,Q^2)$ with
other models.}

\figure{Fig. 4b. Same as Fig. 4a, but for $F_2^n(x,Q^2)$.}

\figure{Fig. 5a. The calculated $[F_2^p(x)-F_2^n(x)]_{val}$
(dashed line), $[F_2^p(x)-F_2^n(x)]_{sea}$ (dot-dashed line)
 and $[F_2^p(x)-F_2^n(x)]_{val+sea}$ (solid line), data are
taken from [6].}

\figure{Fig. 5b. Same as Fig. 5a, but for $[F_2^p(x)-F_2^n(x)]/x$.}

\figure{Fig. 6a. Comparison of our $F_2^p(x)-F_2^n(x)$
with other models (dotted line: [42] and
dot-dashed line: [43]).}

\figure{Fig. 6b. Comparison of $F_2^n(x)/F_2^p(x)$
with data and other models. Solid line (short dashed line):
this paper with $\xi$=0.85 (1.15), dotted line: [42],
dot-dashed line: [43] and dashed line: SU(6) limit.}

\figure{Fig. 7a. $\Delta u_v(x,Q^2)$ calculated from
the model at $Q^2$=0.81, 4.0 and 15.0 (GeV/c)$^2$.}

\figure{Fig. 7b. Same as Fig. 7a, but for $\Delta
d_v(x,Q^2)$.}

\figure{Fig. 8a. The calculated $[xg_1^p(x,Q^2)]_{val}$ at
$Q^2$=0.81 (GeV/c)$^2$ and QCD evolved result at $Q^2$=10.7
(GeV/c)$^2$, data is taken from [12]. The dashed curve is
given by SST model [40].}

\figure{Fig. 8b. Same as Fig. 8a, but for $[xg_1^n(x,Q^2)]_{val}$,
data is taken from [16].}

\figure{Fig. 9a. The calculated $[xg_1^p(x,Q^2)]_{val}$
(dotted curve), $[xg_1^p(x,Q^2)]_{sea}$ (dot-dashed
curve), and $[xg_1^p(x,Q^2)]_{val+sea}$ (solid curve,
data is taken from Ref. [16]).}

\figure{Fig. 9b. Same as Fig. 9a, but for
neutron, data is taken from [16].}

\figure{Fig. 10a. Same as Fig. 9b, but
for $g_1^n(x,Q^2)$.}

\figure{Fig. 10b. Comparison of $[xg_1^d(x,Q^2)]$ with
data [16].}

\figure{Fig. 11a. The calculated $g_2^p(x,Q^2)$ and
$g_2^n(x,Q^2)$ given in this paper (valence contribution
only).}

\figure{Fig. 11b. Comparison of our $g_2^p(x,Q^2)$
(solid line) and those given by other models. Dot-dashed
line: MIT bag model, dashed line: [40].}

\figure{Fig. 12a. The calculated
$g_2(x,Q^2)$ (solid line), $g_2^{WW}(x,Q^2)$ (dotted line)
and ${\bar g}_2(x,Q^2)$
(dot-dashed line) for the proton.}

\figure{Fig. 12b. Same as Fig. 12a, but for the neutron.}

\begin{table}
\caption{Comparison of the calculated moments of unpolarized
quark distribution functions and Gottfried sum rule given by
different models with experiments.}
\begin{tabular}{rclcrclcrclcrcl}
&Quantity       &~~~~&
&Data$^a$       &~~~~&
&F.G. [42]  &~~~~&
&I.K. [43]  &~~~~&
&This paper\\
\tableline
&$\int_0^1u_v(x)dx$&~~~~&&2.04$\pm$0.14 &~~~~&&$-$&~~~~&&$-$&~~~~&&1.999$^*$ \\
&$\int_0^1d_v(x)dx$&~~~~&&1.07$\pm$0.20 &~~~~&&$-$&~~~~&&$-$&~~~~&&0.986 \\
&$\int_0^1xu_v(x)dx$&~~~~&&0.275$\pm$0.011&~~~~&&0.276&~~~~&&0.225&~~~~&&0.275\\
&$\int_0^1xd_v(x)dx$&~~~~&&0.116$\pm$0.017&~~~~&&0.114&~~~~&&0.109&~~~~&&0.123\\
&$\int_0^1xq_s(x)dx$&~~~~&&0.074$\pm$0.011&~~~~&&0.062&~~~~&&0.085&~~~~&&0.079\\
&$[{\rm I}_{\rm Gott.}]_v$&~~~~&&0.333$^b$&~~~~&&$-$&~~~~&&$-$&~~~~&&0.340\\
&$[{\rm I}_{\rm Gott.}]_s$&~~~~&&$-$&~~~~&&$-$&~~~~&&$-$&~~~~&&$-$0.092\\
&${\rm I}_{\rm Gott.}$&~~~~&&0.240$\pm$0.016$^c$&~~~~&&$-$&~~~~&&$-$
&~~~~&&0.248\\
\end{tabular}
\bigskip
a) Data are taken from [1] except ${\rm I}_{\rm Gott.}$.

b) Here $\int_0^1[u_v(x)-d_v(x)]dx=1$ has been used.
Taking the data of $\int_0^1u_v(x)dx$ and $\int_0^1d_v(x)dx$
one obtain $[{\rm I}_{\rm Gott.}]_v$=0.323$\pm$0.081

c) Ref. [6].

*) Input.

\end{table}

\begin{table}
\caption{ Comparison of the calculated moments of polarized
quark distribution functions and hyperon $\beta$-decay
coupling constants given by different models with experiments.}
\begin{tabular}{rclcrclcrclcrcl}
&Quantity       &~~~~&
&Data$^a$       &~~~~&
&MIT [34]  &~~~~&
&SST [40]  &~~~~&
&This paper\\
\tableline
&${\Delta}u_v$&~~~~&&$-$&~~~~&&0.840&~~~~&&0.951&~~~~&&1.0003\\
&${\Delta}d_v$&~~~~&&$-$&~~~~&&$-$0.210&~~~~&&$-$0.202&~~~~&&$-$0.2217\\
&${\Delta}u_s+{\Delta}{\bar
u}_s$&~~~~&&$-$&~~~~&&0.0&~~~~&&$-$&~~~~&&$-$0.229\\
&${\Delta}d_s+{\Delta}{\bar
d}_s$&~~~~&&$-$&~~~~&&0.0&~~~~&&$-$&~~~~&&$-$0.257\\
&${\Delta}s_s+{\Delta}{\bar s}_s$&~~~~&&$-$0.150$\pm$0.090$^b$&
{}~~~~&&0.0&~~~~&&$-$&~~~~&&$-$0.150$^*$\\
&F/D&~~~~&&0.58$\pm$0.02$^c$&~~~~&&0.667&~~~~&&0.703&~~~~&&0.584\\
&   &~~~~&&0.575$\pm$0.016$^d$&~~~~&&     &~~~~&&     &~~~~&&     \\
&$(g_A/g_V)_{n\rightarrow p}$&~~~~&&1.2573$\pm$0.0028&~~~~&&1.050&~~~~&&1.153
&~~~~&&1.250\\
&$(g_A/g_V)_{\Lambda\rightarrow p}$&~~~~&&0.718$\pm$0.015&~~~~&&0.630&~~~~&&
0.701&~~~~&&0.723\\
&$(g_A/g_V)_{\Sigma^-\rightarrow n}$&~~~~&&0.340$\pm$0.017&~~~~&&0.210&~~~~&&
0.250&~~~~&&0.329\\
&$(g_A/g_V)_{\Xi^-\rightarrow \Lambda}$&~~~~&&0.25$\pm$0.05&~~~~&&0.210
&~~~~&&0.202&~~~~&&0.198\\
\end{tabular}
\bigskip
a) Data are taken from [72] except ${\Delta}s_s+{\Delta}{\bar s}_s$
and F/D.

b) Ref. [64].

c) Ref. [67]

d) Ref. [71]

*) Input

\end{table}

\begin{table}
\caption{Comparison of the calculated moments of $g_1^{p,n}$
and $g_2^{p,n}$ given by different models with experiments.}

\begin{tabular}{rclcrclcrclcrcl}
&Quantity  &~~~~&
&Data      &~~~~&
&MIT [34]  &~~~~&
&SST [40]  &~~~~&
&This paper\\
\tableline
&$\int [g_1^p(x)]_vdx$&~~~~&&$-$&~~~~&&0.175&~~~~&&0.220&~~~~&&0.1974\\
&$\int [g_1^n(x)]_vdx$&~~~~&&$-$&~~~~&&0.0  &~~~~&& $-$ &~~~~&&0.0106\\
&$\int [g_1^p(x)]_sdx$&~~~~&&$-$&~~~~&&0.0  &~~~~&& $-$ &~~~~&&$-$0.0714\\
&$\int [g_1^n(x)]_sdx$&~~~~&&$-$&~~~~&&0.0  &~~~~&& $-$ &~~~~&&$-$0.0756\\
&$\int g_1^p(x)dx$&~~~~&&0.126$\pm$0.018$^a$&~~~~&&0.175&~~~~&&0.220
&~~~~&&0.126\\
&$\int g_1^n(x)dx$&~~~~&&$-$0.08$\pm$0.06$^b$&~~~~&&0.0&~~~~&&$-$&~~~~&&
$-$0.065\\
&$\int [g_2^p(x)]_vdx$&~~~~&&$-$&~~~~&&0.0038&~~~~&&$-$&~~~~&&$-$0.0016\\
&$\int [g_2^n(x)]_vdx$&~~~~&&$-$&~~~~&&$-$   &~~~~&&$-$&~~~~&&$-$0.0047\\
\end{tabular}
\bigskip
a) Data is taken from [12].

b) Data is taken from [24].

\end{table}


\begin{references}

\bibitem{sloan88} T. Sloan, G. Smadja, R. Voss, Phys. Rep. $\bf 162$,
45 (1988)

\bibitem{mishra89} S. R. Mishra, F. Sciulli, Ann. Rev. Nucl.
Part. Sci. $\bf 39$, 259 (1989)

\bibitem{roberts91} R. G. Roberts and M. R. Whalley, J. Phys.
$\bf G17$, D1 (1991)

\bibitem{owens92} J. F. Owens, W. K. Tung, Ann. Rev. Nucl. Part. Sci.
$\bf 42$, 291 (1992)

\bibitem{milsztajn91} A. Milsztajn et al., Z. Phys. $\bf C49$, 527 (1991)

\bibitem{amaudruz91} Amaudruz et al., Phys. Rev. Lett. $\bf 66$, 2712 (1991)

\bibitem{gottfried67} K. Gottfried, Phys. Rev. Lett. $\bf 18$, 1174 (1967)

\bibitem{abram82} H. Abramowicz et al., Z. Phys. $\bf C15$, 19 (1982)

\bibitem{foudas90} C. Foudas et al., Phys. Rev. Lett. $\bf 64$
1207 (1990)

\bibitem{alguard76} M. J. Alguard et al., Phys. Rev. Lett. $\bf 37$,
1261 (1976); ibid, $\bf 41$, 70 (1978)

\bibitem{baum80} G. Baum et al.,Phys. Rev. Lett. $\bf 45$,
2000 (1980); ibid, $\bf 51$, 1135 (1983)

\bibitem{ashman89} J. Ashman et al., Phys. Lett. $\bf B206$, 364 (1988);
Nucl. Phys. $\bf B328$, 1 (1989)

\bibitem{hughes83} V. W. Hughes and J. Kuti, Ann. Rev. Nucl. Part. Sci.
$\bf 33$, 611 (1983)

\bibitem{ellis74} J. Ellis, R. L. Jaffe, Phys. Rev. $\bf D9$, 1444 (1974);
$\bf D10$, 1669 (1974)

\bibitem{bjorken66} J. D. Bjorken, Phys. Rev. $\bf 148$, 1467 (1966);
Phys. Rev. $\bf D1$, 1376 (1970)

\bibitem{adeva93} EMC Collaboration, B. Adeva et al., Phys. Lett. $\bf B302$,
534 (1993)

\bibitem{henley93} E. Henley, Bull. Am. Phys. Soc. $\bf 38$, 972 (1993)

\bibitem{ek93} J. Ellis and M. Karliner, Phys. Lett. $\bf B313$, 131 (1993)

\bibitem{brodsky88} S. J. Brodsky, J. Ellis and M. Karliner Phys. Lett
$\bf B206$, 309 (1988)

\bibitem{altarelli88} G. Altarelli and G. G. Ross, Phys. Lett. $\bf B212$,
391 (1988)

\bibitem{carlitz88} R. D. Carlitz, J. C. Collins and A. H. Mueller,
$\bf B214$, 229 (1988)

\bibitem{efremov88} A. V. Efremov and O. V. Teryaev, Dubna preprint
E2-88-287 (1988)

\bibitem{altarelli89} G. Altarelli and W. J. Stirling, Particle World,
$\bf 1$, 40 (1989)

\bibitem{jaffe90} R. L. Jaffe and A. Manohar, Nucl. Phys. $\bf B337$,
509 (1990)

\bibitem{reya91} E. Reya, Dortmund preprint DO-TH 91/09

\bibitem{griffioen92} K. A. Griffioen, AIP Conference Proc. No.269, 271
(1992)

\bibitem{ukawa93} A. Ukawa, Nucl. Phys. $\bf B$ (Proc. Suppl.) $\bf 30$,
3 (1993)

\bibitem{altmeyer93} R. Altmeyer, M. Gockeler, R. Horsley,
E. Laermann and G. Schierholz, Nucl. Phys. $\bf B$ (Proc. Suppl.) $\bf 30$,
483 (1993)

\bibitem{dong93} S. J. Dong and Keh-Fei Liu, Nucl. Phys. $\bf B$
(Proc. Suppl.) $\bf 30$, 487 (1993)

\bibitem{jaffe75} R. L. Jaffe, Phys. Rev. $\bf D11$, 1953 (1975)

\bibitem{hughes77} R. J. Hughes, Phys. Rev. $\bf D16$, 622 (1977);\
J. A. Bartelski, Phys. Rev. $\bf D20$, 1229 (1979)

\bibitem{bell78} J. S. Bell and A. J. G. Hey, Phys. Lett. $\bf 74B$,
77 (1978);\ J. S. Bell, A. C. Davis and J. Rafelski, Phys. Lett.
$\bf 78B$, 67 (1978)

\bibitem{celenza83} L. S. Celenza and C. M. Shakin, Phys. Rev.
$\bf C27$, 1561 (1983); $\bf C39$, 2477 (E) (1989)

\bibitem{jaffe91} R. L. Jaffe and Xiangdong Ji, Phys. Rev. $\bf D43$,
726 (1991)

\bibitem{wang83} X. M. Wang, X. Song, P. C. Yin, Hadron Journal, $\bf 6$,
985 (1983);\ X. M. Wang, Phys. Lett. $\bf 140B$, 413 (1984)

\bibitem{benesh87} C. J. Benesh and G. A. Miller, Phys. Rev. $\bf D36$,
1344 (1987); C. J. Benesh and G. A. Miller, Phys. Rev. $\bf D38$,
48 (1988)

\bibitem{peierls57} R. E. Peierls and J. Yoccoz, Proc. Phys. Soc. $\bf A70$,
381 (1957); R. E. Peierls and D. J. Thouless, Nucl. Phys. $\bf 38$,
154 (1962)

\bibitem{rujula75} A. De Rujula, H. Georgi and S. L. Glashow, Phys. Rev.
$\bf D12$, 147 (1975)

\bibitem{meyer91} H. Meyer, P. J. Mulders, Nucl. Phys. $\bf A528$, 589 (1991)

\bibitem{schreiber91} A. W. Schreiber, A. I. Signal and A. W. Thomas,
Phys. Rev. $\bf D44$, 2653 (1991)

\bibitem{bate92} M. R. Bate, A. I. Signal, J. Phys. $\bf G18$, 1875 (1992)

\bibitem{bickerstaff90} R. P. Bickerstaff, J. L. Londergan, Phys. Rev.
$\bf 42$, 3621 (1990)

\bibitem{traini92} M. Traini, L. Conci and U. Moschella, U. Trento-Louvain
Preprint (1992)

\bibitem{isgur78} N. Isgur and G. Karl, Phys. Rev. $\bf D18$, 4187 (1978);\
$\bf D20$, 2653 (1979)

\bibitem{song92} X. Song, J. S. McCarthy, Phys. Rev. $\bf C46$, 1077 (1992)

\bibitem{altarelli77} G. Altarelli, G. Parisi, Nucl. Phys. $\bf B126$,
298 (1977)

\bibitem{shen83} Shen Qi-xing, Wu Chi-min, Lu Jian-xian and Zhao
Pei-ying, Physica Energiae Fortis et Physica Nuclearis, $\bf 7$, 170 (1983)

\bibitem{song89} X. Song, J. Du, Phys. Rev. $\bf D40$, 2177 (1989)

\bibitem{burkhardt70} H. Burkhardt and W. N. Cottingham, Ann. Phys. (N. Y.)
$\bf 56$, 453 (1970)

\bibitem{feynman77} R. P. Feynman and R. D. Field, Phys. Rev. $\bf D15$,
2590 (1977)

\bibitem{ross79} D. A. Ross and C. T. Sachrajda, Nucl. Phys. $\bf B149$,
497 (1979)

\bibitem{donoghue77} J. F. Donoghue and E. Golowich, Phys. Rev. $\bf D15$,
3421 (1977)

\bibitem{he84} H. He, X. Zhang and Y. Zhuo, Chinese
Physics, $\bf 4$, 365 (1984)

\bibitem{sullivan72} J. D. Sullivan, Phys. Rev. $\bf D5$, 1732 (1972)

\bibitem{henley90} E. M. Henley and G. A. Miller, Phys. Lett. $\bf B251$,
453 (1990)

\bibitem{stern91} J. Stern and G. Clement, Phys. Lett. $\bf B264$, 426 (1991)

\bibitem{signal91} A. Signal, A. W. Schreiber and A. W. Thomas, Mod. Phys.
Lett.
$\bf A6$, 271 (1991)

\bibitem{kumano91} S. Kumano and J. T. Londergan, Phys. Rev. $\bf D44$,
717 (1991)

\bibitem{anselmino92} A. Anselmino, V. Barone, F. Caruso and E. Predazzi,
Z. Phys. $\bf C55$, 97 (1992)

\bibitem{sw92} X. Song and H. J. Weber, Preprint, INPP-UVA-92-17
(unpublished)

\bibitem{weber93} H. J. Weber, Preprint, INPP-UVA-93-5

\bibitem{schuryak89} E. V. Schuryak and J. L. Rosner, Phys. Lett.
$\bf B218$, 72 (1989) and reference there in

\bibitem{feynman72} R. P. Feynman, Photon-Hadron Interactions,
W. A. Benjamin Inc. (1972)

\bibitem{ahrens87} L. Ahrens et al., Phys. Rev. $\bf D35$, 785 (1987)

\bibitem{close90} F. E. Close, Nucl. Phys. $\bf A508$, 413 (1990)

\bibitem{ratcliffe87} P. G. Ratcliffe, Phys. Lett. $\bf B192$, 180 (1987)

\bibitem{hsueh88} S. Y. Hsueh et al., Phys. Rev. $\bf D38$, 2056 (1988)

\bibitem{dorokhov90} A. E. Dorokhov and N. I. Kochlev, Mod. Phys. Lett.
$\bf A5$, 55 (1990)

\bibitem{wandzura77} S. Wandzura and F. Wilczek, Phys. Lett. $\bf 82B$,
195 (1977)

\bibitem{beyer87} M. Beyer and H. J. Weber, Phys. Rev. $\bf C35$, 14 (1987)

\bibitem{close93} F. E. Close and R. G. Roberts, RAL-93-040

\bibitem{pdg92} Particle Data Group, Phys. Rev. $\bf D45$, No.11 (1992)

\end{references}
\end{document}